\documentclass{WileyMSP-template}
\usepackage{comment}
\usepackage{multicol}
\usepackage{xcolor}
\usepackage{mhchem}
\usepackage{parskip}
\usepackage{amssymb}
\usepackage[utf8]{inputenc}
\DeclareUnicodeCharacter{2212}{-}
\DeclareUnicodeCharacter{2009}{=}
\newcommand{\lSAW}{{\lambda_{\mathrm{SAW}}}}		
\newcommand{\vSAW}{v_\mathrm{SAW}}			
\newcommand{\ktwo}{k^2_\mathrm{eff}}
\newcommand{\Sone}{S_\mathrm{11}}	
\newcommand{\Stwo}{S_\mathrm{21}}
\newcommand{\Hfx}{\ce{Al_{1-x}Hf_{x}N}}
\newcommand{\Hf}[2]{\ce{Al_{#2}Hf_{#1}N}}
\newcommand{\sap}{\ce{Al_2O_3}}
\begin{document}

\pagestyle{fancy}

\title{$\Hfx$ ~Thin Films with Enhanced Piezoelectric Responses for GHz Surface Acoustic Wave Devices}

\maketitle


\author{Laura I. Wagner*},
\author{Verena Streibel*},
\author{Esperanza Luna},
\author{Katarina S. Flashar},
\author{Valentin Walbrunn},
\author{Walid Anders},
\author{Nicole Volkmer},
\author{Doreen Steffen},
\author{Katrin Morgenroth}
\author{Frans Munnik},
\author{Tsedenia A. Zewdie},
\author{Saswati Santra},
\author{Jutta Schwarzkopf},
\author{Ian D. Sharp*},
\author{Mingyun Yuan*}


\begin{affiliations}
Dr. Laura I. Wagner, Dr. Verena Streibel, Katarina S. Flashar, Valentin Walbrunn, Tsedenia A. Zewdie, Dr. Saswati Santra, Prof. Dr. Ian D. Sharp\\
Walter Schottky Institute, Technical University of Munich, 85748 Garching, Germany\\
Physics Department, TUM School of Natural Sciences, Technical University of Munich, 85748 Garching, Germany\\
laura.wagner@wsi.tum.de\\
verena.streibel@tum.de\\
sharp@wsi.tum.de

Dr. Esperanza Luna, Walid Anders, Nicole Volkmer, Doreen Steffen, Katrin Morgenroth, Dr. Mingyun Yuan\\
Paul-Drude-Institut f\"ur Festk\"orperelektronik, Leibniz Institut im Forschungsverbund Berlin e.V., 10117 Berlin, Germany\\
yuan@pdi-berlin.de

Dr. Frans Munnik\\
Institute of Ion Beam Physics and Materials Research, Helmholtz-Zentrum Dresden-Rossendorf (HZDR), 01328 Dresden, Germany

Dr. Jutta Schwarzkopf\\
Leibniz-Institut f\"ur Kristallz\"uchtung (IKZ), 12489 Berlin, Germany\\

\end{affiliations}



\today


\begin{abstract}
Ternary compounds obtained by alloying wurtzite AlN with transition metals have emerged as promising materials with significantly enhanced piezoelectric characteristics relative to binary AlN. The increased electromechanical coupling in these compounds boosts the performance of high-frequency acoustic devices. To date, progress has largely focused on aluminum scandium nitride (\ce{Al_{1-x}Sc_xN}), which, while widely deployed, relies on scandium, a costly and geopolitically critical element that is challenging to refine. Here, we investigate aluminum hafnium nitride (\Hfx) as a more economical and accessible alternative to \ce{Al_{1-x}Sc_xN}. Using reactive co-sputtering on both Si and $\alpha$-\ce{Al2O3} substrates, we demonstrate wurtzite \ce{Al_{1-x}Hf_{x}N} thin films ($x \leq 0.17$) with strong $c$-axis texture and nearly isotropic lattice expansion upon Hf incorporation. X-ray absorption spectroscopy indicates cross-gap hybridization between N 2$p$ and Hf 5$d$ states, which can enhance the Born effective charge and, thereby, the piezoelectric response. The incorporation of Hf leads to a multifold enhancement in the piezoelectric coefficient, $d_{33}$, relative to AlN, despite increasing structural disorder in these \Hfx ~thin films. Building on this increased electromechanical coupling, we present functional $\Hfx$ GHz surface acoustic wave (SAW) resonators that exhibit enhanced performance, as well as efficient excitation of bulk acoustic waves with low propagation losses. These results establish $\Hfx$ as a promising platform for next-generation high-frequency electromechanical devices, with prospects for further piezoelectric enhancements through improved epitaxy.       

\end{abstract}


\section{Introduction}
Over the past decade, ternary compounds obtained by alloying wurtzite AlN with transition metals have emerged as a rapidly developing class of materials with promising properties for a range of next-generation electronic and electromechanical devices, enabled by the stable hexagonal wurtzite structure. Since the 2009 discovery that incorporation of Sc into AlN can significantly enhance the piezoelectric strain coefficient, $d_{33}$ \cite{Akiyama2009,Akiyama_review2009}, \ce{Al_{1-x}Sc_xN} thin films have been intensively investigated for applications in radio-frequency (RF) acoustic-wave devices in the GHz range \cite{Hashimoto2012,Wang2014,MOREIRA201123,Wang2018,Yuan_2024}, high-power electronics \cite{Hardy2017,Frei2019,krause_alscngan_2023,nguyen_lattice-matched_2024}, and, more recently, as ferroelectric layers \cite{Fichtner2019,Ryoo.2025, wang_ScAlN_Mo_2023}. The combination of enhanced piezoelectricity with the inherently high thermal conductivity of AlN also makes these materials attractive for use in microelectromechanical systems (MEMS) \cite{wolff2022,Sui.2022,wu2023}. Among these application areas, surface acoustic wave (SAW) devices have garnered particular attention, motivated by the industrial demand for high-frequency, large-bandwidth components in 5G networks and beyond. In this context, key telecommunication elements, such as filters and oscillators, rely on the efficient generation and control of SAWs, which are piezoelectrically-driven waves confined to and propagating along the surface of an elastic solid.

Despite more than fifteen years of ternary aluminum nitride development, \ce{Al_{1-x}Sc_xN} remains the dominant compound for high-performance piezoelectric devices. At the same time, exploring aluminum transition metal nitrides beyond \ce{Al_{1-x}Sc_xN} ~offers advantages not only in terms of material availability but also in enabling different bonding and hybridization mechanisms that can retain and enhance the piezoelectric response. In this regard, a key design principle for enhancing the piezoelectric response in wurtzite ternary nitride compounds is the incorporation of transition metals that distort the lattice, as often reflected by changes in the $c/a$ lattice constant ratio \cite{Farrer_ScGaN2002,Momida.2018,noor-a-alam_ferroelectricity_2019,ambacher_review_2023,Startt.2023}. While Sc incorporation in AlN is the most thoroughly studied example of this effect, recent theoretical and experimental reports have identified several promising alternative elements, including Hf, Zr, Ti \cite{Startt.2023, Iwazaki.2015}, Mo \cite{Feng.2022}, B \cite{Hayden.2021, Calderon.2025}, and Y \cite{Zukauskaite2012,Leone2023}. In particular, theoretical studies suggest that Hf incorporation into wurtzite AlN could produce even larger enhancements than Sc \cite{Startt.2023,wang_piezoelectric_2024}. Although Hf- and Zr-based nitrides are typically stable in cubic structures \cite{GU201559,Sirotti.2024,Wagner.2024b, Streibel.2024}, phase energetics calculations suggest improved wurtzite phase stability for $\Hfx$ compared to \ce{Al_{1-x}Sc_xN} \cite{Lind2013}. This prediction suggests the possibility of incorporating higher concentrations of Hf while maintaining the non-centrosymmetric phase, thereby further enhancing the piezoelectric properties. In addition, Hf incorporation is expected to modify the structural and electronic properties of AlN in ways that are distinct from Sc, with its non-isovalent configuration and bonding characteristics compared to Al, offering additional control over the resulting physical properties. Together, these factors suggest that $\Hfx$ could be a promising alternative to \ce{Al_{1-x}Sc_xN} for next-generation piezoelectric devices. Although few experimental demonstrations of Al-Hf-N compounds exist, earlier studies have mainly focused on rocksalt structures \cite{Selvakumar.2015, Rauchenwald.2020} or the co-doping of wurtzite AlN with Mg and Hf \cite{Nguyen.2017, Yokoyama.2015, Bradford2025}. In contrast, recent reports on the ferroelectricity and increased piezoelectricity of $\Hfx$ \cite{Bernstein.2025}, its conduction mechanisms \cite{Bernstein2026}, and structural anomalies such as deviations in the $c/a$ lattice constant ratio compared to \ce{Al_{1-x}Sc_xN} \cite{Wiebke2026} strongly motivate a dedicated exploration and improved understanding of this emerging materials system.

In this work, we synthesized \ce{Al_{1-x}Hf_{x}N} thin films with $0 \le x \le 0.17$ by pulsed DC magnetron co-sputtering and investigated their crystalline and electronic structures, as well as their electromechanical properties. The layers were deposited on both n$^+$-Si(100), of relevance for future CMOS integration, and (0001)-oriented $c$-plane $\alpha$-\ce{Al2O3} (sapphire, for brevity denoted as $c$-\ce{Al2O3} for the remainder of the work), which provides a suitable surface for epitaxial growth and acoustic wave device demonstration. Structural characterization combining X-ray diffraction (XRD) and transmission electron microscopy (TEM) reveals wurtzite phase formation on both substrates, with nearly isotropic lattice expansion upon Hf incorporation. On $c$-\ce{Al2O3}, the sputtered films are partially epitaxial, exhibiting pronounced $c$-axis and in-plane orientational order. Complementary X-ray absorption spectroscopy (XAS) indicates that Hf incorporation preserves the local bonding environment of wurtzite AlN, while introducing enhanced cross-gap hybridization that can increase the Born effective charge and, hence, the piezoelectric response. Consistent with this observation and the prior work of Bernstein et al. \cite{Bernstein.2025}, piezoforce microscopy (PFM) and complementary double-beam laser interferometry (DBLI) reveal enhanced piezoelectric properties in our Hf-containing films. Building on these experimental indications, we demonstrate that \ce{Al_{1-x}Hf_{x}N} on $c$-\ce{Al2O3} functions as a powerful platform for GHz SAW resonators, with \ce{Al_{0.94}Hf_{0.06}N} exhibiting enhanced SAW generation efficiency compared to AlN. In addition, we observe the excitation and detection of GHz bulk acoustic waves (BAWs) with ultra-low propagation loss. Notably, these promising characteristics are achieved with sputtered \ce{Al_{1-x}Hf_{x}N} thin films having a reduced structural order compared to the sputtered binary AlN reference films, highlighting the potential for future optimization of this emerging materials system  . Together, our findings provide key insights into the structural, electronic, and electromechanical properties of $\Hfx$, demonstrating that this promising new materials system can serve as a powerful platform for RF piezoelectric systems, ranging from SAW resonators to MEMS devices.

\section{Results and Discussion} 

\subsection{Synthesis of \ce{Al_{1-x}Hf_xN} thin films}

\begin{figure}[h!]
    \centering
    \includegraphics{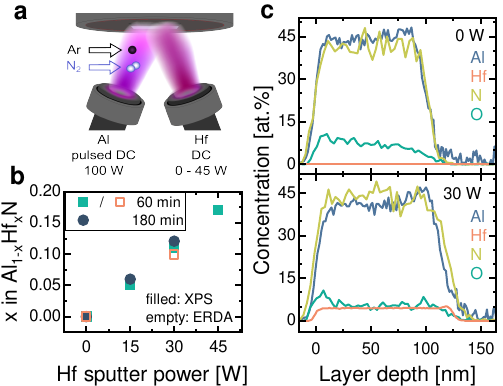}
    \caption{Synthesis of $\Hfx$ thin films and analysis of their elemental compositions. \textbf{(a)} Schematic illustration of reactive magnetron co-sputtering of $\Hfx$ using two metallic targets. \textbf{(b)} Hf cation fraction as a function of Hf sputter power in $\Hfx$ thin films deposited for 60 min (squares) and 180 min (circles), as determined by XPS (filled symbols) and ERDA (empty squares). \textbf{(c)} ERDA elemental composition depth profiles measured on samples sputtered with 0\,W (upper panel) and 30\,W (lower panel) applied to the Hf sputter target.}
    \label{fig:1chemcomp}
\end{figure}

$\Hfx$ thin films with controlled Hf content were deposited by reactive magnetron co-sputtering (Figure \ref{fig:1chemcomp}a) on (100)-oriented n$^+$-Si and $c$-\ce{Al2O3} to enable systematic investigation of their structural, electronic, and electromechanical properties (see Experimental Section). In brief, a fixed pulsed DC bias was applied to a metallic Al target in an Ar:\ce{N2} process gas, while Hf was incorporated by co-sputtering from a separate metallic Hf target operated in DC mode, with the applied power varied to control the Hf content. Film thicknesses were defined by the deposition time, which was 60\,min for an initial set of thinner samples and 180\,min for a subsequent set of thicker films that were better suited for SAW-device testing. The resulting thicknesses were determined via spectroscopic ellipsometry (SE), with the samples deposited for 60\,min ranging from 110 -- 140\,nm and those deposited for 180\,min ranging from 270 -- 350\,nm, depending on the Hf sputter power (see Table S1).

The chemical compositions of the sputtered films were determined using a combination of X-ray photoelectron spectroscopy (XPS) and elastic recoil detection analysis (ERDA). Representative XPS survey spectra for films deposited on n$^+$-Si are shown in Figure S1, with corresponding high-resolution scans of O 1s, N 1s, Al 2p, and Hf 4f core level regions given in Figure S2. We note that exposure of the samples to ambient conditions results in formation of a thin native oxide layer, such that surface-sensitive XPS overestimates the oxygen and underestimates the nitrogen content in the films. Nevertheless, analysis of the metal core level intensity ratios enables determination of the Hf cation fraction, $x$ in \ce{Al_{1-x}Hf_{x}N}, as a function of the Hf sputter power. As shown in Figure \ref{fig:1chemcomp}b, the Hf cation fraction increases linearly with Hf sputter power, reaching a maximum value of $x=0.17$ for a sputter power of 45\,W. In addition, we find that the surface cation composition as a function of Hf sputter power is independent of film thickness, highlighting the reproducibility and compositional control enabled by the reactive co-sputter deposition process.

To assess the bulk compositions of the films, ERDA depth profiling was performed on two selected samples. As indicated in Figure \ref{fig:1chemcomp}b, the average cation fractions determined by ERDA are in excellent agreement with those obtained by XPS. The corresponding ERDA elemental depth profiles (Figure \ref{fig:1chemcomp}c) exhibit constant Al and N concentrations of 40 -- 45\,at.\% for both the 0 W (binary AlN) and 30 W Hf samples, with the latter containing a constant 4.8\,at.\% Hf ($x = 0.1$) throughout the layer. We observe, however, a slight Hf enrichment at the film-substrate interface, which becomes visible in the zoomed-in profile in Figure S3. This unintentional interfacial feature will be discussed in more detail in the TEM analysis below. Finally, oxygen impurity concentrations decrease with film depth, reaching a bulk content of 5 -- 8\,at.\%. Such a surface enrichment is commonly observed in (oxy)nitride thin films \cite{Wagner.2024b, Streibel.2024} and is likely a consequence of ambient oxidation of the surface and near-surface grain boundaries \cite{patidar_deposition_2024}. Overall, the combined XPS and ERDA results confirm nearly stoichiometric cation-to-nitrogen ratios in the binary AlN and ternary \Hfx ~thin films. 

\subsection{Crystallographic and microstructural analysis of \ce{Al_{1-x}Hf_xN}}\label{SecXRD}

Grazing incidence XRD (GIXRD) confirms that all films are in the wurtzite AlN structure \cite{Ott.1924}, regardless of Hf content, film thickness, or substrate type (Figure S4). No reflections from other AlN polymorphs or rocksalt HfN are observed, indicating that the films are phase-pure to within the detection limit. Increasing Hf concentrations lead to a shift of all reflections to lower $2\theta$ angles, consistent with an expansion of the unit cell volume in the ternary compounds.

\begin{figure}[h!]
    \centering
    \includegraphics{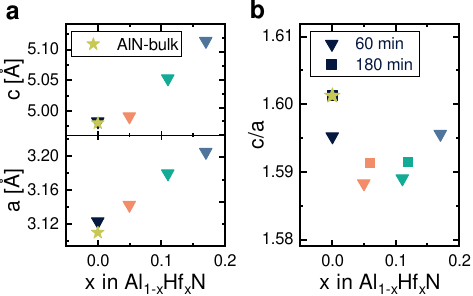}
    \caption{Lattice parameters of \Hfx. \textbf{(a)} Lattice constants, $c$ (upper panel) and $a$ (lower panel), of $\Hfx$ thin films deposited for 60\,min on Si as a function of Hf cation fraction. For reference, the corresponding lattice constants of bulk AlN (stars) are also provided \cite{Ott.1924}. \textbf{(b)} Lattice parameter ratio $c/a$ for the thin films deposited for 60 min on Si (inverted triangles) and for thicker films deposited for 180 min on $c$-\ce{Al2O3} (squares).}
    \label{fig:lattpar}
\end{figure}

Previous theoretical studies have reported conflicting findings regarding the dependence of lattice parameters on Hf content within \Hfx. While Wang and co-workers predicted a decreasing $c/a$ ratio, even at low Hf concentrations ($x=0.12$) \cite{wang_piezoelectric_2024}, Startt et al. suggested a nearly constant value for Hf cation fractions below $x=0.2$ \cite{Startt.2023}. Here, high-resolution XRD (HRXRD) out-of-plane $2\theta/\omega$ scans, combined with in-plane $2\theta _\chi / \phi$ diffraction measurements (Figure S5) reveal that both $a$ and $c$ increase with Hf cation fraction (Figure \ref{fig:lattpar}a). As a consequence, the $c/a$ ratio remains close to 1.6 across the investigated composition range and is independent of film thickness (Figure \ref{fig:lattpar}b). This nearly isotropic expansion is thus consistent with the theoretical calculations by Startt et al. \cite{Startt.2023}, as well as with recent experimental reports by Bernstein et al. \cite{Bernstein.2025, Bernstein2026} and Liebscher et al. \cite{Wiebke2026}. However, the observed trend differs from that of \ce{Al_{1-x}Sc_xN}, in which Sc incorporation induces an anisotropic distortion, even at low Sc contents, resulting in a $c/a$ ratio that decreases to below 1.5 in \ce{Al_{0.75}Sc_{0.25}N} \cite{Dargis.2020, Dinh.2025}. Considering that this reduced $c/a$ ratio in \ce{Al_{1-x}Sc_xN} has been linked to its increased piezoelectric response compared to AlN \cite{Momida.2018}, as well as the emergence of its ferroelectricity \cite{Zhang_Sc_content2013}, the apparent absence of a similar anisotropic distortion in $\Hfx$ raises important questions about how similar property enhancements might be achieved in this compound, as discussed further in Section \ref{PFM}.

Residual stress, which can influence the performance characteristics of thin film electromechanical devices, was estimated for the AlN films using the method detailed in Ref. \cite{Bartasyte_residue_stress_2012}, taking lattice and stiffness constants for bulk AlN from Refs.~\cite{Ott.1924} and \cite{McNeil_AlN_1993}, respectively. The in-plane biaxial stress is estimated to be -0.6~GPa (60 min deposition) and -1.0~GPa (180 min deposition), indicating compressive stress. Since no powder or bulk references are available for \Hfx ~($x>0$), the residual stress in Hf-containing films cannot be reliably estimated using this method.

\begin{figure}[h]
    \centering
    \includegraphics{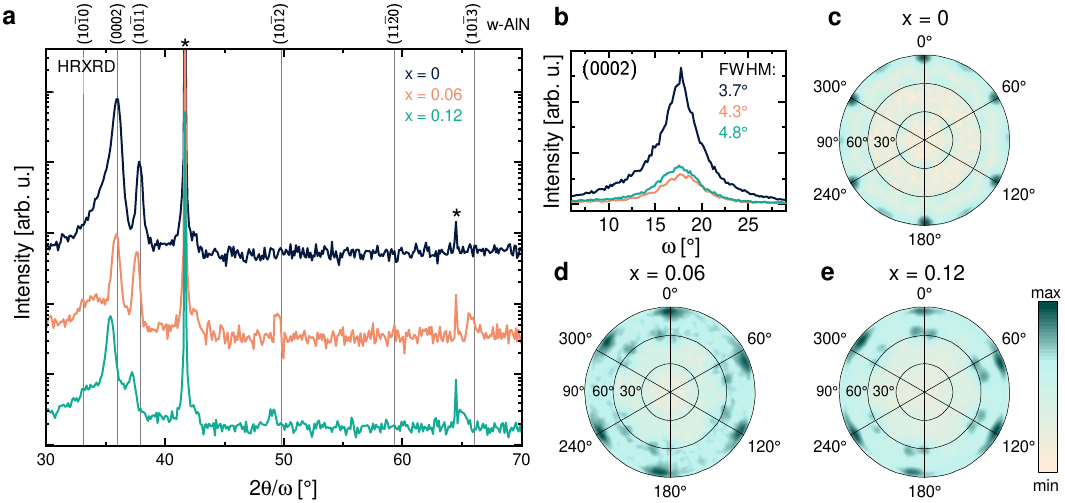}
    \caption{Structures of sputtered \ce{Al_{1-x}Hf_{x}N} thin films. \textbf{(a)} HRXRD (out-of-plane) $2\theta/\omega$ scans of \ce{Al_{1-x}Hf_{x}N} thin films on $c$-\ce{Al2O3} displayed on a logarithmic intensity scale. The vertical gray reference lines indicate the diffraction pattern of wurtzite AlN, labeled with their Miller-Bravais indices \cite{Ott.1924}. Stars mark the $(0006)$ and $(0009)$ peaks of \ce{Al2O3} \cite{Kortan_ScN_Al2O3,Inaba.2013}.
    \textbf{(b)} Rocking curve scans on the $(0002)$ reflection of wurtzite \Hfx. In-plane pole figures on a logarithmic intensity scale of the $(10\overline{1}0)$ diffraction peak of \textbf{(c)} AlN, \textbf{(d)} \ce{Al_{0.94}Hf_{0.06}N}, and \textbf{(e)} \ce{Al_{0.88}Hf_{0.12}N}.}
    \label{fig:2Structure}
\end{figure}

Having characterized the general crystallographic changes induced by Hf incorporation, we now turn to an analysis of the structural properties of thicker ($\approx$ 300\,nm) \ce{Al_{1-x}Hf_{x}N} films on $c$-\ce{Al2O3} substrates, which are of greatest relevance for assessing SAW performance characteristics. In particular, we comparatively analyze three \ce{Al_{1-x}Hf_xN} films, with $x =$ 0 (binary AlN), 0.06, and 0.12. Across all compositions, out-of-plane HRXRD scans ($2\theta/\omega$) exhibit sharp substrate reflections at 41.7° and 64.5° from the $(0006)$ and $(0009)$ planes of $c$-\ce{Al2O3} \cite{Kortan_ScN_Al2O3,Inaba.2013}, as well as wurtzite $\Hfx$ reflections, with a dominant $(0002)$ peak near 36° and a weaker $(10\overline{1}1)$ feature at 38° (Figure \ref{fig:2Structure}a). We note that the (0009) reflection of \ce{Al2O3} is formally forbidden but may appear in diffraction patterns of \ce{Al2O3} single crystals due to multiple beam diffraction, known as the \textit{Renninger effect} \cite{Renninger.1937}. Additionally, weak $(10\overline{1}2)$ and $(10\overline{1}3)$ reflections are visible in the Hf-containing samples. Compared to the predicted powder diffraction intensity distribution \cite{Ott.1924}, in which the $(10\overline{1}0)$ peak is the most intense, the strength of the $(0002)$ peak suggests considerable thin film texture, with growth primarily oriented along the $c$-axis. In summary, the reduced intensity ratio between the $(0002)$ and $(10\overline{1}1)$ reflections, the reduced $(0002)$ intensity, and the appearance of additional low-intensity reflections from $(10\overline{1}2)$ and $(10\overline{1}3)$ planes indicate increased grain misorientation with Hf incorporation. 

X-ray rocking-curve (XRC) measurements on the $(0002)$ reflection provide additional insight into the mosaicity and tilt disorder in these films. As shown in Figure \ref{fig:2Structure}b, the AlN reference film exhibits a full width at half maximum (FWHM) of 3.7° with a narrow coherent peak that is commonly associated with low-density, ordered misfit dislocations in epitaxial layers \cite{Kaganer2009}. The FWHM increases to 4.3° and 4.8° for the Hf-containing films with $x=0.06$ and $x=0.12$, respectively. Although these values exceed those from previous reports of sputtered AlN \cite{Bernstein.2025, Dargis.2020}, a preferred $c$-axis orientation of the $\Hfx$ thin films is still maintained. While for binary AlN, the intensity and sharpness of the (0002) reflection are widely used as indicators of piezoelectric quality, additional composition-dependent geometric and electronic factors play key roles in defining the piezoelectric response in ternary nitrides, as will be discussed in Section~\ref{PFM}. 

Pole figure measurements further confirm the strong $c$-axis orientation of the films, while also revealing the impact of Hf incorporation on their in-plane orientation relationships. 
For the AlN reference film, the $m$-plane $(10\overline{1}0)$ pole figure exhibits six sharp maxima at $\Psi = 80-90$°, separated by 60° in $\phi$, consistent with the six-fold symmetry of wurtzite AlN, confirming a strong $c$-axis fiber texture and in-plane orientation relationship (Figure \ref{fig:2Structure}c). In addition, weaker secondary maxima, rotated by 30°, indicate partial in-plane biaxial alignment. Corresponding pole figures from \ce{Al_{1-x}Hf_{x}N} with $x=0.06$ (Figure \ref{fig:2Structure}d) and $x=0.12$ (Figure \ref{fig:2Structure}e) reveal that the six-fold symmetry is retained, but the diffraction maxima broaden and become less evenly spaced in $\phi$, indicating reduced in-plane alignment and increased mosaicity. In addition to the primary six-fold pattern, the Hf-containing films also contain weaker peak pairs at $\Psi$ of $\sim$60°, consistent with the angle between $(10\overline{1}3)$ and $(10\overline{1}0)$, indicating partial twinning. Comparing the pole figure symmetries with the substrate orientation (Figure S6) reveals that all films possess an in-plane twist of approximately 30° with respect to $c$-\ce{Al2O3}, which is a well-known growth mode for III-nitrides on $c$-\ce{Al2O3} \cite{Soomro.2016, Sun.1994}. Taken together with supporting pole figure measurements on $(0002)$ and $(10\overline{1}1)$ diffraction planes (Figure S6), these results confirm that the \ce{Al_{1-x}Hf_{x}N} films are partially epitaxial, with strong $c$-axis orientation but notable azimuthal broadening and secondary orientations. 

\begin{figure}[h!]
    \centering
    \includegraphics{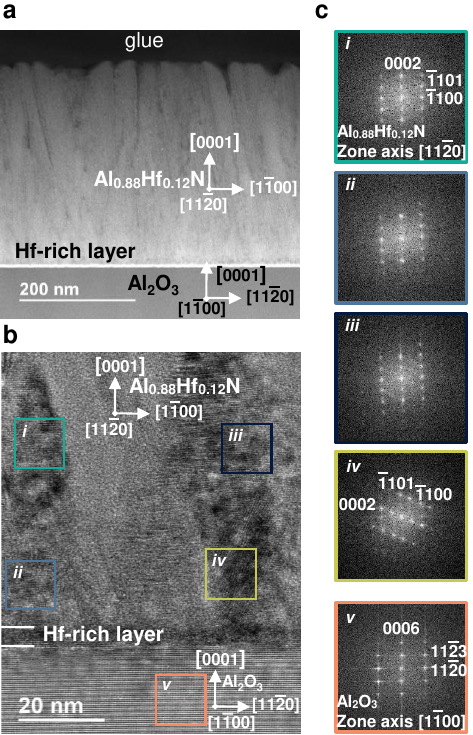}
    \caption{Microstructure of sputtered \ce{Al_{1-x}Hf_{x}N} thin films. Representative \textbf{(a)} HAADF-STEM and \textbf{(b)} HRTEM images of 340~nm thick {\Hf{0.12}{0.88}} films grown on $c$-\ce{Al2O3}, along with \textbf{(c)}, FFT images from the selected areas indicated by the corresponding boxes {\it i-v}.}
    \label{fig:TEM}
\end{figure}

High-resolution TEM (HRTEM) of the \ce{Al_{0.88}Hf_{0.12}N} film on $c$-\ce{Al2O3} reveals the microscopic origin of the structural disorder observed by XRD. Figure \ref{fig:TEM}a shows a representative high-angle annular dark-field scanning TEM (HAADF-STEM) image taken along the \ce{Al2O3} [$1\bar100$] zone axis. The bulk of the $\Hfx$ layer, well above the interface, consists of coalesced columnar grains with average diameters of $\sim$30\,nm. Although growth along the $c$-axis dominates within this bulk region, domains with preferred growth on semipolar planes, possibly $\{10\bar{1}3\}$ and $\{10\bar{1}1\}$, are also observed. Closer inspection of the \Hfx/\ce{Al2O3} interfacial region reveals a finer and more disordered grain structure. In addition, we observe bright contrast in HAADF-STEM micrographs at the film/substrate interface (Figure ~\ref{fig:TEM}a), which is assigned to the presence of an unintentional Hf-rich interlayer based on its high Z-contrast and further supported by the interfacial Hf enrichment observed in the ERDA profile (Figure S3). The formation of this layer disrupts the ideal epitaxial relationship at the interface and likely contributes to the structural imperfections observed by XRD. While it is highly likely that this layer influences grain orientation, its effect on lattice strain and wurtzite phase stability is difficult to quantify. This difficulty is primarily due to the limited availability of structural reference data for wurtzite \Hfx, which makes it challenging to disentangle strain contributions arising from Hf incorporation within the lattice from those potentially induced by the Hf-rich interlayer.

HRTEM phase-contrast imaging along the [$1\bar100$] zone axis of \ce{Al_2O_3} provides insight into the impact of the Hf-rich interlayer on local crystal orientations. Fast Fourier transform (FFT) analysis of selected regions, indicated by the boxes numbered {\it i-v} within Figure ~\ref{fig:TEM}b, reveal grains with different orientations near the interface (Figure ~\ref{fig:TEM}c). Despite the presence of the Hf-rich interlayer, certain regions retain full epitaxial alignment between \Hf{0.12}{0.88} (box {\it i}) and \ce{Al_2O_3} (box {\it v}), while other regions display quasi-$c$-axis orientations ({\it ii} and {\it iii}). Such tilts are consistent with the mosaic spread indicated by the $(0002)$ rocking curves discussed above. In addition, domains oriented with $\{10\bar11\}$ planes parallel to the surface are observed (box {\it iv}). Additional HRTEM analysis along the [$11\bar20$] zone axis  of \sap~(Figure S7) confirms the existence of domains that adopt semipolar $\{10\bar13\}$ orientations, reflecting the rhombohedral symmetry of \ce{Al2O3}. Such alignment likely leads to the tilted domains observed in Figure \ref{fig:TEM}a and is consistent with the semipolar orientations identified through XRD and planar diffraction analysis. 

Overall, the combined XRD and TEM analysis demonstrates that reactive co-sputtering enables deposition of highly-oriented wurtzite $\Hfx$ thin films on $c$-\ce{Al2O3}, with clear evidence for $c$-axis alignment and epitaxial domains. However, the presence of structural disorder, as given by the mosaicity and misoriented grains, suggests opportunities for future growth optimization. In particular, the thin Hf-rich interlayer at the film-substrate interface, which likely forms during target conditioning and contributes to disruption of the epitaxial relationship. Subsequent growth experiments show that introducing a thin AlN buffer layer prior to Hf target conditioning substantially reduces the (0002) rocking-curve FWHM on both Si and c-Al\textsubscript{2}O\textsubscript{3} substrates (Figure S8), demonstrating that improved structural quality can be achieved by suppressing formation of this parasitic interlayer.  Additional strategies, including the use of alternative substrates, such as hexagonal SiC or high-temperature post-deposition annealing, may provide further opportunities for growth optimization. Since such optimization strategies were not applied to the sample series used for the measurements presented below, the piezoelectric properties and SAW efficiencies should be viewed as conservative lower bounds, with considerable future improvements expected as growth processes are refined. 

\subsection{Electronic structure and piezoelectric properties} \label{PFM}
A key finding from the structural analysis presented above is that incorporation of Hf into AlN leads to an approximately isotropic lattice expansion, with the $c/a$ ratio remaining nearly constant across the studied composition range. This behavior stands in stark contrast to other ternary compounds such as \ce{Al_{1-x}Sc_xN}, where the reduction of $c/a$ upon Sc incorporation has been linked to its enhanced piezoelectricity and the emergence of ferroelectricity \cite{Momida.2018,ambacher_review_2023,Zhang_Sc_content2013},  with $c/a$ identified as a viable proxy for computational screening of these important functional properties. In this context, recent reports of enhanced piezoelectricity and ferroelectric switching in $\Hfx$ appear surprising, since the absence of a significant change in $c/a$ would seem to preclude such effects \cite{Startt.2023,Bernstein.2025}. However, not only geometric effects but also the magnitude of the Born effective charge affect the calculated piezoelectric constants in III-nitrides \cite{noor-a-alam_ferroelectricity_2019,Bernardini1997,Wang-polarization2021}. The Born effective charge describes how the movement of an ion within a crystal induces an electronic polarization, effectively capturing the strength of the coupling between ionic motion and the resulting electric field. Most remarkably, increased piezoelectricity via Sc incorporation in III-nitrides was predicted based on this mechanism \cite{Farrer_ScGaN2002}, seven years before the first experimental reports. Unlike Sc, Hf is not isovalent with Al, suggesting that differences in the outer shell configuration could lead to modified orbital hybridization and electronic polarization effects that are fundamentally different from the geometric effects mainly discussed for \ce{Al_{1-x}Sc_xN} and related isovalent substitutions of Al \cite{Momida.2018,Zhang_Sc_content2013}. 

\begin{figure}
    \centering
    \includegraphics{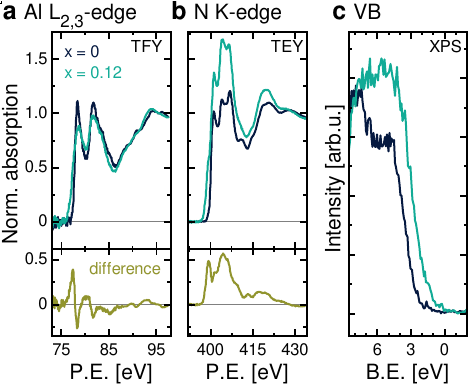}
    \caption{Electronic structure of \Hfx. Normalized X-ray absorption spectra of AlN and \ce{Al_{0.88}Hf_{0.12}N} thin films at the \textbf{(a)} Al L$_{2,3}$- and \textbf{(b)} N K-edge, collected via total fluorescence yield (TFY) and total electron yield (TEY), respectively, as a function of photon energy (P.E.). The corresponding differences between the \ce{Al_{0.88}Hf_{0.12}N} and the binary AlN reference spectra are given in the lower panels. \textbf{(c)} Valence band spectra measured by XPS as a function of binding energy (B.E.)}
    \label{fig:XAFS}
\end{figure}

Based on the considerations above, we hypothesize that electromechanical changes to $\Hfx$ within the investigated composition range may be primarily electronic rather than structural in origin. To test this possibility, we performed XAS at the Al L$_{2,3}$-edge and N K-edge, which provide element- and site-specific probes of the conduction band (CB) electronic structure, bonding, and local coordination environment. Figures \ref{fig:XAFS}a and b show a comparison of XAS spectra from a binary AlN reference film and from \ce{Al_{0.88}Hf_{0.12}N}, normalized to each edge step height, along with corresponding difference spectra to highlight changes arising from Hf incorporation. The fine structures are similar for both samples and are characteristic of wurtzite AlN \cite{Serin.1998, Mizoguchi.2003}, consistent with the presence of the tetrahedral bonding symmetry. However, we observe distinct changes to the spectra upon Hf incorporation, with the Al L$_{2,3}$-edge broadening and shifting by -0.9 eV for \ce{Al_{0.88}Hf_{0.12}N} compared to AlN, as indicated by the derivative-like shape of the difference spectrum (Figure \ref{fig:XAFS}a, lower panel). In addition, we find a pronounced increase of the near-edge absorption strength in the N K-edge, as indicated by the predominantly positive component of the difference spectrum and an additional pre-edge peak arising at approx. 399 eV (Figure \ref{fig:XAFS}b, lower panel). Importantly, since both films contain comparable N concentrations and the spectra are normalized to the step height, such differences cannot arise from elemental ratio differences alone. Rather, these changes suggest that Hf incorporation leads to an increased contribution of N 2p states within the conduction band, which is consistent with enhanced cross-gap N 2p--Hf 5d hybridization. In line with this interpretation, spectroscopic ellipsometry (Figure S9) of our \ce{Al_{1-x}Hf_{x}N} films indicates an increase of the high-frequency dielectric constant, $\varepsilon_{\infty}$, which would also be expected from increased electronic polarizability induced by enhanced cross-gap hybridization.

Generally, cross-gap hybridization would be expected to lower the energy of the conduction band minimum (CBM) and increase the Born effective charge. Indeed, similar hybridization effects and corresponding increases in the Born effective charge have been implicated in enhanced piezoelectric and ferroelectric properties in a range of oxide perovskites \cite{Cohen.1992, Zhong_polarization1994, Ghosez.1998}, thallium halides \cite{Du.2010}, and \ce{LaWN3} \cite{Fang.2017}. In case of \ce{Al_{1-x}Sc_{x}N}, however, such hybridization effects have not yet been reported based on XAS measurements, likely because the strong overlap of the N K- and Sc L$_{2,3}$-edges poses a significant challenge for reliable spectroscopic analysis and interpretation.  

While the XAS results presented above provide strong indications for enhanced cross-gap hybridization, the non-isovalent nature of Hf could also lead to significant electronic doping effects. In particular, substitution of \ce{Al^{3+}} with \ce{Hf^{4+}} should result in introduction of donor-like states, local defect complexes, or increased free charge carrier concentrations. To investigate whether this is the case, we measured the valence band (VB) region via XPS. As shown in Figure \ref{fig:XAFS}c, increasing Hf content leads to a decrease in the energetic separation between the VB maximum (VBM) and the Fermi level ($E_\mathrm{F}$), as well as an increase in the relative photoemission intensity near the VBM, suggesting the introduction of additional occupied electronic states. Nevertheless, the Fermi level remains well within the bandgap, indicating that the films remain non-degenerate semiconductors. In addition, complementary spectroscopic ellipsometry measurements indicate vanishingly small extinction coefficients across the sub-bandgap range, regardless of Hf content (not shown). Thus, the lack of metallic-like states near $E_\mathrm{F}$ and the absence of a Drude-like optical response confirm that our $\Hfx$ films retain their dielectric character across the investigated composition range, despite the non-isovalency between Hf and Al. This behavior is consistent with prior studies by Bernstein et al., which associate non-isovalent Hf incorporation with charge-compensating defect states and localized electronic states at defect sites \cite{Bernstein.2025, Bernstein2026}. In principle, such doping-related effects can also modify the dielectric response and screen polarization fields, thereby influencing the effective piezoelectric coupling. While the present results are consistent with N 2p-Hf 5d hybridization playing a dominant role in the enhanced piezoelectric response within the investigated composition range, doping-related effects may become increasingly relevant at higher Hf concentrations and warrant systematic future investigations. 

To directly establish whether the observed electronic structure changes in $\Hfx$ lead to enhanced piezoelectric properties, as well as to assess the suitability of this material for SAW resonators, we measured the out-of-plane piezoelectric strain coefficient, $d_{33}$, using both PFM and DBLI. For these measurements, it was necessary to use films grown on $n^+$-Si(100) (60 min deposition time), which provides an electrically conductive substrate for reliable bottom contact formation. However, it is important to note that such films are characterized by increased structural disorder compared to those grown on $c$-\ce{Al2O3} (Figure S10), including a significantly larger mosaic spread of the desired $c$-axis orientation. As such, quantified values of $d_{33}$ were obtained from structurally non-optimized material, suggesting possible improvements in  more ideally epitaxial films.

\begin{figure}[h!]
    \centering
     \includegraphics{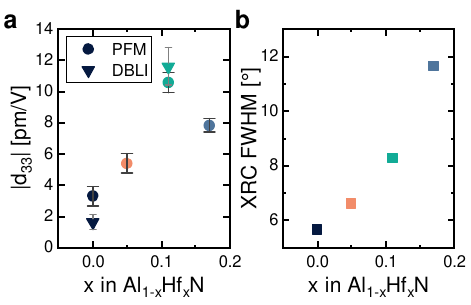}
    \caption{Piezoelectric properties of \Hfx ~thin films. 
    \textbf{(a)} Piezoelectric coefficient $\vert d_{33}\vert$ extracted from the spatially averaged PFM lock-in amplitude as a function of Hf concentration, together with representative DBLI measurements obtained for pure AlN and \Hf{0.11}{0.89} ~samples. The local PFM and macroscopic DBLI measurements are in good agreement, both indicating a significant enhancement of the piezoelectric response upon Hf incorporation. Error bars for the PFM data represent the standard deviation of the PFM response across the scanned area, while those for the DBLI data represent the standard deviation of the $ d_{33}$ values obtained over the bias-dependent measurements (Figure S14). The values are extracted from the PFM maps (Figure S12) following compensation for tip-sample potential differences measured by KPFM (Figure S11).  
    \textbf{(b)} Corresponding FWHM values from XRCs of the (0002) reflection, indicating that piezoelectric enhancements occur despite increasing structural disorder within the films and, thus, should be taken as lower limits.}
    \label{fig:PFM}
\end{figure}

We summarize the $\vert d_{33}\vert$ values converted from the PFM lock-in amplitudes in Figure~\ref{fig:PFM}a. Complete KPFM and PFM maps used for calculation of $ d_{33}$ for all compositions are provided in Figures S11 and S12, respectively. Here, incorporation of moderate concentrations of Hf leads to a systematic enhancement of $ \vert d_{33}\vert$, reaching a value of 10.6 pm/V for \ce{Al_{0.89}Hf_{0.11}N} that is above a factor of three larger than that of the AlN reference film, 3.3 pm/V. 
Importantly, this increase occurs despite the progressively increasing structural disorder with Hf content, as evidenced by the broadening of the $(0002)$ rocking curves on Si (Figures~\ref{fig:PFM}b and S13). Nevertheless, for the highest Hf content film (\ce{Al_{0.83}Hf_{0.17}N}), $ d_{33}$ decreases, which we ascribe to the considerable structural disorder within this film rather than a fundamental characteristic of $\Hfx$ since the material remains below the wurtzite-to-rocksalt phase transition. 

Bulk AlN typically has a $ d_{33}$ of $\sim$5.5\,pm/V \cite{Guy1999,Bernardini2002}. While single-crystal $\Hfx$ has been theoretically predicted to have a $ d_{33}$ of $\sim$10\,pm/V for $x=0.12$~\cite{wang_piezoelectric_2024} and $\sim$8\,pm/V for $x=0.17$~\cite{Startt.2023},
the values measured in the thin-film configuration are generally lower than those of bulk samples since the transverse $d_{31}$ and the relevant compliance coefficients lead to a negative contribution, yielding a reduced effective $d_{33}$ \cite{Lefki1994}. In addition, we note that determination of the piezoelectric coefficient $d_{33}$ in thin films using PFM remains challenging compared to more quantitative methods such as double‑beam laser interferometry (DBLI), as employed by Bernstein et al. \cite{Bernstein.2025}. Therefore, we performed complementary DBLI measurements on the $x = 0$ and $x = 0.11$ films. These representative DBLI measurements yield values of approximately 2 pm/V for AlN and approximately 11 pm/V for \ce{Al_{0.89}Hf_{0.11}N} (corresponding bias-dependent measurements are provided in Figure S14). Thus, both the local PFM and macroscopic DBLI measurements are in good agreement, indicating a significant enhancement of the piezoelectric response upon Hf incorporation.  These results are consistent with the enhanced piezoelectricity reported previously \cite{Bernstein.2025}, as well as with theoretical predictions \cite{Startt.2023,Bernstein.2025}.  Thus, our results confirm that Hf incorporation can significantly enhance the piezoelectric response of AlN, even in the presence of increasing disorder, offering considerable promise for $\Hfx$ as a functional material for GHz SAW devices.  In this context, a recent study demonstrated that the $d_{33}$ value of \ce{Al_{1-x}Sc_{x}N} can be significantly enhanced through thermal post-processing \cite{mondal2025}, suggesting a promising route to further optimization.

\subsection{Surface acoustic wave and bulk acoustic wave excitation}
\begin{figure}[h!]
    \centering
    \includegraphics{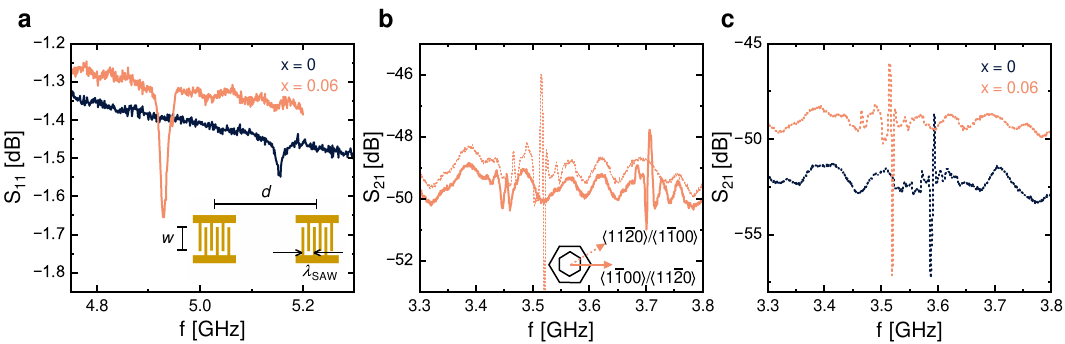}
    \caption{Excitation of SAWs in \Hfx/$c$-\ce{Al2O3}. \textbf{(a)} $\Sone$ for IDTs with $\lSAW=1$ $\mu$m along the \Hfx/\ce{Al2O3} $\langle1\bar100\rangle$/$\langle11\bar20\rangle$ directions. Inset: illustration of a delay line formed by two IDTs. \textbf{(b)} Comparison of SAW resonances detected in \Hf{0.06}{0.94} along \Hfx/\ce{Al2O3} $\langle1\bar100\rangle$/$\langle11\bar20\rangle$ and $\langle11\bar20\rangle$/$\langle1\bar100\rangle$ for $\lSAW=1.5$ $\mu$m, indicating the in-plane anisotropy. The two inequivalent directions are illustrated by the arrows in the inset, which shows a smaller $\Hfx$ ~hexagon overlaid on a larger \sap ~hexagon. \textbf{(c)} $\Stwo$ for delay lines with $\lSAW=1.5$ $\mu$m along \Hfx/\ce{Al2O3} $\langle11\bar20\rangle$/$\langle1\bar100\rangle$. In all cases, the delay distance is $d=200\lSAW$.
    }
    \label{fig:SAW1}
\end{figure}

Building on the enhanced piezoelectric response observed by PFM, we next evaluate whether $\Hfx$ thin films can provide improved performance in high-frequency SAW devices. For this purpose, we fabricated Ti/Al/Ti interdigital transducers (IDTs) on the thicker $\Hfx$ layers, sputtered for 180 min on $c$-\ce{Al2O3}, and measured their resonant acoustic responses as a function of Hf content. As illustrated in the inset of Figure~\ref{fig:SAW1}a, two opposing sets of IDTs were fabricated to form delay line resonators with varying propagation lengths, $d$, while the SAW wavelengths, $\lSAW$, were defined by the fabricated IDT finger periodicity. 
All devices were otherwise designed with identical geometries, comprising the same number of finger pairs, $N$, and beam widths, $w$, to enable comparison across different Hf concentrations.

S-parameter measurements from the fabricated delay line resonators confirm efficient SAW excitation across all investigated compositions. As shown in Figure~\ref{fig:SAW1}a, clear resonances are observed in the input reflection coefficient ($\Sone$) near 5~GHz for SAW propagation ($\lSAW=1$ $\mu$m) along the \Hfx/\ce{Al2O3} $\langle1\bar100\rangle$/$\langle11\bar20\rangle$ directions. Strikingly, the $\Hf{0.06}{0.94}$ film exhibits a stronger $\Sone$ resonance than the binary AlN reference film, despite the broader mosaic spread and greater degree of grain misorientation in the ternary film. This behavior is consistent with the increased piezoelectric coefficient of $\Hfx$ films (Figure \ref{fig:PFM}a) and indicates that incorporation of moderate concentrations of Hf improves the SAW generation efficiency in the GHz range. 

To investigate the influence of in-plane anisotropy on SAW propagation, we fabricated delay lines along two high-symmetry directions offset by 30$^\circ$ in plane. As discussed in Section~\ref{SecXRD}, the $\Hfx$ films adopt a well-defined 30$^\circ$ in-plane rotation with respect to $c$-\ce{Al2O3}, a common feature of epitaxial III-nitrides on $c$-\ce{Al2O3}. This relationship is illustrated in the inset of Figure~\ref{fig:SAW1}b, where the basal planes of \Hfx ~and $c$-\sap~ are represented by the smaller and larger hexagons, respectively. This configuration gives rise to two inequivalent families of SAW propagation directions, indicated by the arrows, namely \Hfx/\sap$\langle11\bar20\rangle$/$\langle1\bar100\rangle$ (dotted line) and $\langle1\bar100\rangle$/$\langle11\bar20\rangle$ (solid line). 

\begin{figure}[h!]
    \centering
    \includegraphics{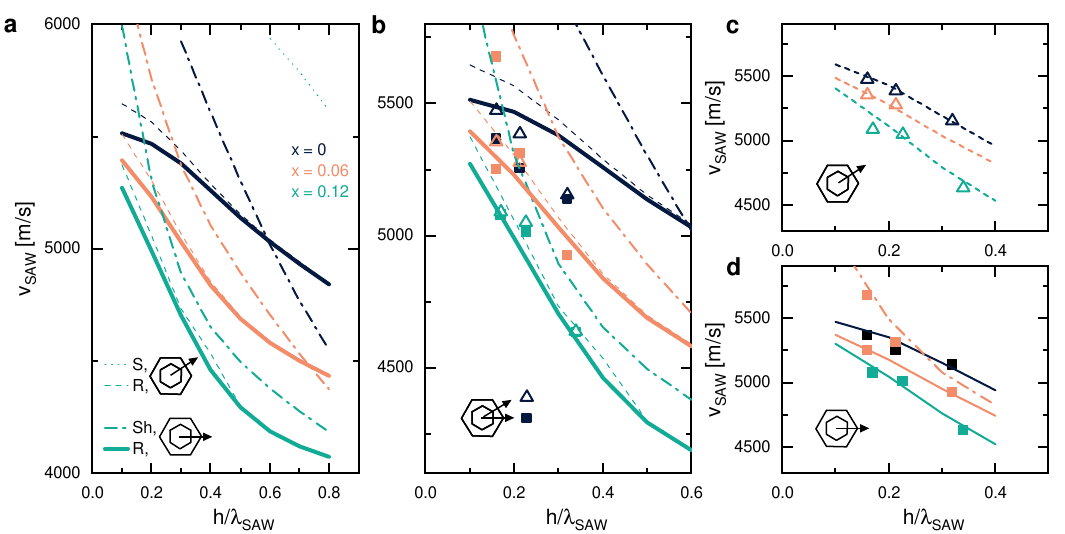}
    \caption{Summary of simulated and measured SAW modes. \textbf{(a)} Calculated SAW mode velocities as a function of the normalized film thickness, Hf content, and propagation direction obtained using the initial material parameters (Table S3) for $x$ up to 0.12. The two directions are illustrated in the inset, which shows a smaller $\Hfx$ ~hexagon overlaid on a larger \sap ~hexagon. R: Rayleigh-type SAW; Sh: shear-type SAW; S: Sezawa SAW. \textbf{(b)} Zoom-in of the simulated dispersion curves from (a) compared to the experimentally-measured resonances. \textbf{(c)} and \textbf{(d)} Experimentally detected resonances compared to the simulation performed with tuned material parameters (Table S4) along \Hfx/\ce{Al2O3} $\langle11\bar20\rangle$/$\langle1\bar100\rangle$ and $\langle1\bar100\rangle$/$\langle11\bar20\rangle$, respectively.
    }
    \label{fig:4SAW}
\end{figure}

Notably, \Hf{0.06}{0.94} exhibits two resonance modes (at 3.46 and 3.71~GHz) for propagation along \sap $\langle11\bar20\rangle$, while only a single mode is detected along the \sap $\langle1\bar100\rangle$ direction (Figure~\ref{fig:SAW1}b). The clear detection of all modes near 3.6 GHz, even for the long delay length of $d=200\lSAW$ ($\lSAW=1.5$~$\mu$m), highlights the efficient piezoelectric coupling in these \Hf{0.06}{0.94} devices. Furthermore, as shown in Figure~\ref{fig:SAW1}c, the forward transmission coefficient $\Stwo$ along \sap  $\langle1\bar100\rangle$ yields a nearly 3-dB enhancement for \Hf{0.06}{0.94} compared to AlN. The observed in-plane anisotropy and the two-mode behavior along \sap $\langle11\bar20\rangle$ reflect both the symmetry of rhombohedral \sap~and, more importantly, the strong in-plane orientation of the \Hf{0.06}{0.94} thin film, as discussed further in connection with Figure~\ref{fig:4SAW}. Extraction of the electromechanical coupling coefficient, $\ktwo$, can be found in Figure~S15 and related discussion in the Supporting Information. Hence, two different approaches both confirm enhanced $\ktwo$ values for the Hf-containing films. For $x=0.12$ the resonance signal returns to a level comparable to AlN (Figure~S16), as the piezoelectric enhancement is offset by the progressive formation of misoriented grains.

To further interpret the experimental SAW results, we used the transfer matrix method to calculate the acoustic dispersion of \ce{Al_{1-x}Hf_{x}N} on $c$-\ce{Al2O3} with non-piezoelectric Ti/Al electrodes for all three compositions and both propagation directions. The stiffness tensor elements, $C_{IJ}$, and piezoelectric stress tensor elements, $e_{iJ}$, were initially obtained from prior density functional theory (DFT) calculations of \Hf{0.12}{0.88} \cite{wang_piezoelectric_2024} and from literature data for AlN \cite{Deger98a,Tsubouchi_IEEESU32_634_85}, as described in the Experimental Section and summarized in Table S3. As shown in Figure~\ref{fig:4SAW}a, with different compositions distinguished by color, the dispersion was evaluated as a function of the film thickness normalized to the SAW wavelength ($h/\lSAW$). For each composition, multiple SAW branches are predicted, and the acoustic velocity decreases with increasing $h/\lSAW$. In addition, the entire dispersion curve shifts downwards with increasing Hf content, as expected for the higher mass density and, possibly, lattice softening of the ternary compound. Along the \Hfx/\sap $\langle1\bar100\rangle$/$\langle11\bar20\rangle$ directions, the first mode corresponds to the fundamental Rayleigh (R) mode (solid line), which is predominantly polarized in the sagittal plane as a coupled shear vertical-longitudinal (SV-L) mode. The second mode is a fundamental shear-type (Sh) SAW (dash-dot line) with a dominant shear-horizontal (SH) component. A crossing of the R- and Sh-branches occurs within the calculated range for $x=0$ and $x=0.06$, where the two modes become degenerate.

Along the 30° in-plane rotated \Hfx/\sap $\langle11\bar20\rangle$/$\langle1\bar100\rangle$ directions, the electromechanical coupling of the Sh-mode remains negligible across the entire range, similar to previous reports on \ce{Al_{1-x}Sc_xN} \cite{Feil2021}. Therefore, only the R-mode is shown for each composition (dashed lines). As $h/\lSAW$ approaches zero, the R-mode propagating along \sap $\langle1\bar100\rangle$ exhibits a higher $\vSAW$ than along \sap $\langle11\bar20\rangle$, reflecting the respective shear-vertical (SV) limits in \sap. With increasing $h/\lSAW$, the difference diminishes and the velocities of the two R-modes converge, while the anisotropy at small $h/\lSAW$ further decreases with increasing Hf content. In addition, for $x=0.12$, $\vSAW$ decreases sufficiently to satisfy the "slow-on-fast" condition when $h/\lSAW>0.6$, leading to the emergence of the second order SV-L mode, commonly referred to as the Sezawa (S) mode (dotted line).

The experimentally observed SAW resonances follow the general trends predicted by our dispersion calculations (Figure~\ref{fig:4SAW}b). Across all compositions and both propagation directions, the SAW velocities obtained from the resonance frequencies ($\vSAW = f_{\mathrm{res}} \lSAW$) decrease with increasing $h/\lSAW$ and with increasing Hf content. The velocity difference between the R-modes along \sap $\langle11\bar20\rangle$ (solid squares) and $\langle1\bar100\rangle$ (empty triangles) is effectively described, and for $x=0.06$ both R-mode and Sh-mode are clearly detected along the $\langle11\bar20\rangle$ direction. Quantitatively, the agreement between experiment and simulation is reasonable for $x=0.12$, but less satisfactory for the other two compositions. However, it is important to note that these simulations were based on a combination of materials parameters from previously reported DFT calculations, as well as literature data for the AlN reference, rather than experimental measurements of our films. Therefore, we refined a restricted set of elastic tensor elements in the simulations (see Experimental Section and Supporting Information), yielding the stiffness and piezoelectric constants given in Table S4. The resulting dispersion relations for the \sap $\langle1\bar100\rangle$ (Figure~\ref{fig:4SAW}c) and $\langle11\bar20\rangle$ (Figure~\ref{fig:4SAW}d) directions yield considerably better agreement with the experimental data. Although the resulting set of refined parameters is not unique, it can provide several insights into the general elastic behavior of \Hfx. The AlN film appears mechanically softer than predicted by the initial $C_{IJ}$ values, while the variation in the diagonal elements of $C_{IJ}$ with composition is smaller than initially assumed. Therefore, the reduction of $\vSAW$ primarily originates from the increased mass density when Al is substituted by Hf. In addition, the off-diagonal shear component $C_{12}$ is larger than initially estimated, resulting in hybridization of the R- and Sh-modes near $h/\lSAW=0.2$ for $x=0.06$, where the crossing evolves into an anti-crossing. Importantly, because the Sh-mode is strongly sensitive to in-plane anisotropy, it can only be detected if both the thin film and the substrate exhibit a clear orientational relationship, both perpendicular to and within the basal plane. In contrast, for fully textured films with randomized in-plane orientations, the Sh resonance would be washed out. The clear observation of the Sh-mode in the \Hf{0.06}{0.94} film thus confirms its high degree of in-plane orientation, consistent with the XRD results (Figure~\ref{fig:2Structure}). 

\begin{figure}[h!]
    \centering
    \includegraphics{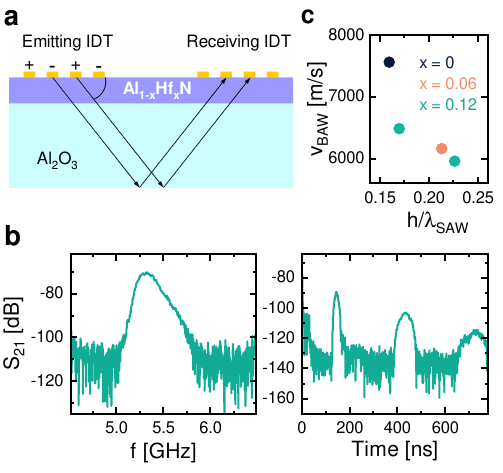}
    \caption{Generation and experimental observation of BAW modes. \textbf{(a)} Schematic of the excitation, reflection, and detection of BAWs propagating with an angle $\alpha$ with respect to the surface. \textbf{(b)} $\Stwo$ spectra of the reflected BAW in the frequency and time domain, excited by and detected in a delay line structure ($d=500\lSAW$) on \Hf{0.12}{0.88} with IDTs of 1.5 $\mu$m periodicity. For the frequency domain, a band-pass filter between 100 and 300~ns was applied to suppress the interference arising from the subsequent echos (Fabry-Perot modes). \textbf{(c)} Extracted $v_\mathrm{BAW}$ as a function of the normalized film thickness and \Hf{x}{1-x} composition.}
    \label{fig:BAW}
\end{figure}

In addition to SAWs, bulk acoustic waves (BAWs) are efficiently excited in the \Hfx/$c$-\ce{Al2O3} heterostructures when the in-plane wavenumber defined by the IDT periodicity is phase-matched to the bulk acoustic modes of the layered system. The resulting waves propagate at an angle $\alpha$, reflect from the polished backside of the \ce{Al2O3} substrate, and are detected at the receiving IDT, as illustrated in Figure \ref{fig:BAW}a. The corresponding frequency- and time-domain $\Stwo$ spectra (Figure \ref{fig:BAW}b), measured on an \Hf{0.12}{0.88} film on $c$-\ce{Al2O3}, exhibit a broad 3 dB bandwidth of 0.2 GHz and multiple well-resolved echoes, indicating low acoustic losses. The BAW velocities, $v_\mathrm{BAW}$, were determined from the device geometry and resonance frequency, $f_\mathrm{res}$, according to: 

\begin{equation}
    \frac{v_\mathrm{BAW}}{f_\mathrm{res}\lSAW}=\cos{\alpha}; \qquad \alpha\sim\tan^{-1}{\frac{h_\mathrm{s}}{d/2}}
\end{equation}

where $h_\mathrm{s}$ is the substrate thickness and $d$ the delay distance. The extracted velocities, plotted as a function of the normalized film thickness in Figure \ref{fig:BAW}c, range from $\sim$7500 m/s for AlN to $\sim$6000 m/s for the Hf-containing films. Furthermore, time-domain analysis of the echoes reveals an ultra-low propagation loss of 0.007~dB$/\mu$m, which is a result of the high acoustic quality of \ce{Al2O3} together with efficient BAW excitation and detection by the $\Hfx$ thin film piezoelectric transducers. Finite element method (FEM) simulations of the acoustic mode profiles (Figure S17) reproduce the angle $\alpha$ of the sample configuration. Apart from the bulk-penetrating BAW modes, the surface-confined SAW mode is also plotted in Figure S17 for comparison. Their respective $f_\mathrm{res}$ is in agreement with our experimental observations. The combination of experimental results and FEM simulations indicates that broadband BAWs can be excited with simple, surface IDTs as opposed to conventional thin-film bulk acoustic resonators (fBARs) involving complex, vertical multilayer fabrication. In general, these experimental and simulation results demonstrate that \Hfx/$c$-\ce{Al2O3} heterostructures support not only efficient SAW generation but also bulk acoustic modes with extremely low propagation losses. 

\section{Conclusion}
In conclusion, we have demonstrated that phase-pure wurtzite $\Hfx$ thin films with tunable composition up to at least $x=0.17$ can be synthesized by reactive co-sputtering. Comprehensive structural characterization reveals that films deposited on both (100) Si and (0001)-oriented $c$-plane $\alpha$-\ce{Al2O3} ($c$-\ce{Al2O3}) are strongly $c$-axis oriented and, on $c$-\ce{Al2O3}, possess a preferred in-plane orientational relationship with the substrate. Still, they contain a relatively broad mosaic spread and grain misorientations that could be improved through future growth optimization. In contrast to the intensively investigated \ce{Al_{1-x}Sc_xN} system, Hf incorporation into AlN leads to an approximately isotropic lattice expansion, with the $c/a$ ratio remaining near 1.6 across the investigated composition range. Complementary XAS measurements reveal enhanced cross-gap $p$-$d$ hybridization, which can increase the Born effective charge and has been invoked to explain the enhanced piezoelectricity in other material classes. Consistent with this mechanism, PFM measurements show that the piezoelectric coefficient, $d_{33}$, triples with Hf incorporation, despite the increasing structural disorder in the Hf-containing films. Together, these findings indicate that the increased piezoelectric response primarily arises from electronic effects, which stands in contrast to the key role of the anisotropic lattice distortion in \ce{Al_{1-x}Sc_xN}. 

Based on these structural and electronic insights, we fabricated GHz SAW resonators and demonstrated that Hf incorporation enhances the electromechanical coupling relative to binary AlN. Detailed analysis of the velocity dispersion of both Rayleigh and shear-type SAWs along two high symmetry crystallographic directions confirms that Hf incorporation preserves the wurtzite structure and epitaxial relationship required for anisotropic acoustic wave propagation. Comparison with numerical calculations based on reported theoretical predictions of the stiffness coefficients provides first insights into the previously unexplored acoustic properties of this new piezoelectric compound. 
In addition, we demonstrate efficient excitation and detection of BAWs with low propagation losses. The proof-of-concept GHz SAW devices based on $\Hfx$ already outperform AlN, despite the reduced structural order of these sputtered films. Subsequent growth experiments using an AlN buffer layer demonstrate that the structural quality can be significantly improved, providing a clear route toward further optimization of piezoelectric properties and SAW efficiencies.  Overall, this work demonstrates that $\Hfx$ is a promising, CMOS-compatible alternative to \ce{Al_{1-x}Sc_xN} for next-generation high-frequency acoustic and MEMS devices.  


\section{Experimental Section}\label{SecExp}
\threesubsection{Thin film deposition}

The $\Hfx$ thin films were deposited using reactive magnetron co-sputtering on (100)-oriented n$^{+}$-Si (Siegert Wafer, As-doped, $\leq$0.005\,$\Omega$ cm, single-side polished) and (0001)-oriented c-plane $\alpha$-\ce{Al2O3} (Siegert Wafer, double-side polished, $c$-\ce{Al2O3}). Prior to deposition, the substrates were consecutively rinsed with DI water, acetone, and isopropanol, followed by drying with pressurized dry \ce{N2}. Film deposition was performed using a Kurt J. Lesker Company Proline PVD 75 sputtering system equipped with a metallic Al target (Kurt J. Lesker Company, Al 99.9995\,\%, 2-inch) and a metallic Hf target (Kurt J. Lesker Company, 99.9\,\% , 2-inch) at a target-to-substrate distance of 17\,cm, and with a base pressure below $5\times10^{-8}$\,Torr. For conditioning, each target was pre-sputtered at 60\,W for 10\,min in 9.5\,mTorr of pure Ar. During this time, a 20\,W RF substrate bias was applied to sputter clean the substrates and remove the native oxide from the Si surfaces. The Ar‑ion RF pre‑conditioning step used prior to deposition is intentionally mild, applied only to gently clean the substrate surface without significantly altering or damaging the surface. Future optimization of the growth process will include exploring high‑temperature $c$-\sap substrate annealing prior to deposition, which is widely used to promote improved surface morphology and may further enhance epitaxial alignment \cite{uehara_IEEE_2002}. 

The $\Hfx$ thin films were sputtered at a substrate setpoint temperature of 800\,°C, a working pressure of 6.6\,mTorr, and in a gas mixture of Ar/\ce{N_2} with a ratio of 12/20. The Al target power was fixed at 100\,W, operated in pulsed DC mode with a frequency of 100\,kHz and a reverse time of 2\,$\mu$s. The Hf target was run in DC sputtering mode with an applied power ranging from 0 -- 45\,W enabling control of the Hf concentration within the films. Depending on the desired film thickness, the deposition time was 60 or 180\,min, after which the films were allowed to naturally cool in vacuum.

\threesubsection{Structural, compositional, optical, and electronic characterization}

All XRD measurements were performed using a Rigaku SmartLab diffractometer equipped with a copper anode and a HyPix-3000 2D detector. High-resolution XRD (HRXRD) was measured using a 2-bounce Ge monochromator and scanning $2\theta/\omega$ after alignment to the lattice normal plane of either the \ce{Al2O3} or Si substrates. Grazing incidence X-ray diffraction (GIXRD) was measured at a fixed incidence angle, $\omega$, of 0.4° while scanning 2$\theta$ from 25° to 70°. In-plane $2\theta _\chi / \phi$ scans on $c$-plane \ce{Al2O3} were measured with $\phi$ aligned to the $(30\overline{3}0)$ \ce{Al2O3} reflection and $\omega$ = 0.3-0.5°.

TEM measurements were performed with a JEOL 2100F microscope operated at 200 kV and equipped with a bright-field (BF) detector, a high-angle annular dark field (HAADF) detector, and a Gatan Ultra Scan 4000 charge coupled device. Cross-sectional TEM specimens were prepared for observation along the orthogonal $[1\bar100]$ \ce{Al2O3} and $[11\bar20]$ \ce{Al2O3} projections using standard mechanical polishing and dimpling, followed by Ar ion milling. Analysis of high-resolution (HR)TEM phase contrast micrographs and diffractograms was performed using the Gatan Inc. Digital Micrograph™ (DM) software and JEMS simulation program \cite{STADELMANN1987131,JEMS}.

XPS was measured using a PHOIBOS 150 NAP hemispherical analyzer (SPECS) paired with a focused, monochromated Al K$_\alpha$ X-ray source (1486.6\,eV; XR 50 MF with $\mu$FOCUS 450, SPECS). During measurements, the base pressure was held below 6×10$^{-9}$\,mbar. Survey scans and high-resolution core-level spectra were recorded with pass energies of 90\,eV and 30\,eV, respectively. Spectral analysis was performed using CasaXPS to fit quasi-Voigt line shapes after applying a Shirley background subtraction.

ERDA was performed at the Ion Beam Center of the Helmholtz–Zentrum Dresden-Rossendorf. These measurements were conducted using a 43\,MeV Cl$^{7+}$ ion beam at a fixed angle of 75° between the sample normal and the incoming beam, along with a scattering angle of 30°. The analyzed area was $\approx$2\,$\times$ 2\,mm$^2$. The recoil atoms and scattered ions were detected with a Bragg ionization chamber, which enabled energy measurement and Z identification of the particles. Analysis of ERDA data was performed using the Windf v9.3 g software \cite{Barradas.1997}.

Variable angle spectroscopic ellipsometry was performed using an M2000 XI-210 ellipsometer by J. A. Woollam Company and the CompleteEASE software to estimate film thicknesses and optical constants. Spectra were collected using light in the range of 210 -- 1690\,nm emitted by a xenon lamp (L2174-01, 75\,W, Hamamatsu) at incident angles between 45° and 75°, as well as in transmission. The data were fitted starting from a generalized oscillator model for AlN available in the CompleEASE software.

XAS measurements at the Al L$_{2,3}$ and N K-edge were carried out at the IOM-CNR BEAR beamline at Elettra. All samples were measured in TFY and TEY mode. Each spectrum was energy-calibrated using the second derivative of the corresponding reference absorption spectrum, as well as intensity-corrected by subtraction of a dark scan value measured immediately before or after each scan. The data were normalized, background corrected, and merged using the Athena software package \cite{Ravel.2005}.

Force microscopy was performed using a Nanosurf FlexAFM atomic force microscopy (AFM) system and Pt-coated, conductive Si cantilevers (Nanosensors PPP-EFM) with a free resonance of 78 kHz. Scans were acquired with a scan time of 3 s per line. The surface potential was first determined by Kelvin probe force microscopy (KPFM) measurements operated in dynamic mode at 78 kHz using the primary lock-in amplifier (Figure S11). A DC tip voltage, dynamically controlled to zero the amplitude of a secondary lock-in operating at 13 kHz, was used to determine the contact potential difference between tip and sample. The reported value corresponds to the average over the scanned region. PFM measurements were subsequently performed in static mode with the primary lock-in amplifier disabled. Off-resonance detection was employed to ensure a more stable response, and the operating frequency was set to 13 kHz. The tip-sample contact potential difference determined  by KPFM was compensated during the PFM measurements to minimize artifacts originating from electrostatic effects \cite{Balke2015}. For each sample, an area of $2\times2$ $\mu$m$^2$ was scanned  (Figure~S11). The area-averaged PFM lock-in amplitude was used to determine the corresponding  $ d_{33}$ value, while the RMS variation across the scanned region was used to define the error bars. The value of $ d_{33}$ was calculated from the lock-in amplitude $V_{\mathrm {lock-in}}$ by using the calibrated off-resonance AFM sensitivity $\sigma=382$~nm/V and a lock-in offset $\delta_{\mathrm {lock-in}}=10$~$\mu$V, yielding $d_{33}=(V_{\mathrm {lock-in}}-\delta_{\mathrm {lock-in}})\cdot\sigma/V_{\mathrm {ac}}$.

DBLI was performed using an aixACCT TF 3000 analyzer for electroceramic materials with a DBLI extension. The backsides of the n$^+$-Si substrates were mechanically polished using diamond paper to obtain an optically reflective surface and were uniformly coated with a stack of Ti/Al/Ti (10/30/10 nm) as bottom contacts. Circular top electrodes with a diameter of 400 µm were then deposited on the top surface of the film using the same metallization. The He-Ne laser measurement beam was first reflected from the top contact and later at the bottom contact. The center intensity of the interference pattern between the measurement beam and a reference beam was detected by a photodiode. Square pulses of $V_\mathrm{ac} = 20$ V at 300 Hz were applied across the top and bottom contacts for a lock-in detection, while the DC bias was cycled between $V_\mathrm{dc} = -2.5$ and $3.5$~V.

\threesubsection{Surface acoustic wave excitation and numerical calculation}

IDTs were patterned using either laser lithography (Heidelberg Instruments DWL66+) or electron-beam lithography (Raith150 Two), followed by a standard lift-off process. A Ti/Al/Ti metal stack was deposited using electron-beam evaporation to form
IDTs on the $\Hfx$ layers. The standard thicknesses of these metal layers were 10/50/10 nm, though thinner layers of 10/30/10 nm were used for the devices presented in Figure~\ref{fig:SAW1}b,c. The SAW wavelengths $\lSAW$ were determined by the IDT finger periodicity of 2.0, 1.5, and 1.0 \ce{\mu}m. 
Each IDT comprised 100 finger pairs with a beam width of 60 $\mu$m. RF characterization was performed using an RF probe station (S\"uss MicroTec PM5), following calibration with 50 $\Omega$ standards to eliminate effects of the connecting cables to the RF probes ($\vert Z\vert$-Probe Z20-X-SG/GS-250). S-parameters between 2 to 6 GHz were measured with a network analyzer (Hewlett Packard 8720D).

Calculation of the SAW velocities was performed using a C++ numerical simulation software based on the differential equations for propagation of acoustic modes, using the transfer matrix procedure to include multiple layers. Additional information about the numerical technique can be found in "Appendix. Numerical Simulations" of Ref.~\cite{PVS120}.
During the calculation, the total metal thickness was fixed at 70 nm, ~\Hfx ~film thicknesses were given by the experimentally determined values, and $\lSAW$ was varied according to the step size of $h/\lSAW=0.1$. The initial parameters used in the calculation are summarized in Table S3. The stiffness $C_{IJ}$ and piezoelectric $e_{33}$ constants for $x=0$ and $x=0.12$ were obtained from Ref.~\cite{wang_piezoelectric_2024}. For $x=0.06$ the parameters were not readily available and were interpolated from the other two compositions. Missing elements were taken from bulk AlN data \cite{Deger98a,Tsubouchi_IEEESU32_634_85}. Static dielectric constants $\epsilon_{ij}$ were experimentally obtained by measuring the IDT capacitance and were assumed to be isotropic. The film density $\rho$ was computed by scaling the value of AlN, considering the ratio in weighted atomic weight due to Hf substitution as well as unit-cell volume (the change in the latter is minimal). Tuned parameters used to obtain better agreement with the experimental SAW resonances (Figure \ref{fig:4SAW}c, d) are summarized in Table S4. Stabilization conditions for hexagonal structures were maintained with these values, and unlisted parameters remained unchanged from their initial values.

FEM simulations of SAW and BAW profiles were performed using the programs GetDP \cite{getdp,getdp-ieee1998} and Gmsh \cite{gmsh}.

\medskip
\textbf{Acknowledgements} \par 
We thank Oliver Brandt, Oliver Ambacher, Paulo V. Santos, Duc Van Dinh, Christoph Pratsch and Lutz Geelhaar for valuable discussions, as well as Paulo V. Santos for providing the SAW simulations framework. In addition, we thank Alexander Tselev for the advice on PFM. I.D.S. and M.Y. acknowledge financial support from Deutsche Forschungsgemeinschaft (DFG, German Research Foundation) SPP2744 under Project Number 563184308. This project has received funding from the European Research Council (ERC) under the European Union's Horizon 2020 research and innovation program (grant agreement no. 864234), from the DFG under Germany's Excellence Strategy – EXC 2089/2 – 390776260, and from the German Ministry of Research, Technology and Space within the project SINATRA:CO2UPLED (project number 033RC034). V.S. acknowledges support from the Bavarian Academy of Sciences and Humanities. J.S. acknowledges funding from European Regional Development Fund (ERDF) (Project No. 1.8/15). Parts of this research were carried out by F.M. at the Ion Beam Center at the Helmholtz-Zentrum Dresden-Rossendorf e. V., a member of the Helmholtz Association. We acknowledge Elettra Sincrotrone Trieste for providing access to its synchrotron radiation facilities, and we thank Angelo Giglia and Nicola Mahne for assistance in using beamline BEAR (Proposal ID 20245364).

\textbf{Supporting Information} Supporting Information is available from the Wiley Online Library.

%
\bibliographystyle{MSP}

\medskip


\



\end{document}


\pagestyle{fancy}

\title{Supporting Information: $\Hfx$~Thin Films with Enhanced Piezoelectric Responses for GHz Surface Acoustic Wave Devices}

\maketitle

\author{Laura I. Wagner*},
\author{Verena Streibel*},
\author{Esperanza Luna},
\author{Katarina S. Flashar},
\author{Valentin Walbrunn},
\author{Walid Anders},
\author{Nicole Volkmer},
\author{Doreen Steffen},
\author{Katrin Morgenroth}
\author{Frans Munnik},
\author{Tsedenia A. Zewdie},
\author{Saswati Santra},
\author{Jutta Schwarzkopf},
\author{Ian D. Sharp*},
\author{Mingyun Yuan*}

\begin{affiliations}
Dr. Laura I. Wagner, Dr. Verena Streibel, Katarina S. Flashar, Valentin Walbrunn, Tsedenia A. Zewdie, Dr. Saswati Santra, Prof. Dr. Ian D. Sharp\\
Walter Schottky Institute, Technical University of Munich, 85748 Garching, Germany\\
Physics Department, TUM School of Natural Sciences, Technical University of Munich, 85748 Garching, Germany\\
laura.wagner@wsi.tum.de\\
verena.streibel@tum.de\\
sharp@wsi.tum.de

Dr. Esperanza Luna, Walid Anders, Nicole Volkmer, Doreen Steffen, Katrin Morgenroth, Dr. Mingyun Yuan\\
Paul-Drude-Institut f\"ur Festk\"orperelektronik, Leibniz Institut im Forschungsverbund Berlin e.V., 10117 Berlin, Germany\\
yuan@pdi-berlin.de

Dr. Frans Munnik\\
Institute of Ion Beam Physics and Materials Research, Helmholtz-Zentrum Dresden-Rossendorf (HZDR), 01328 Dresden, Germany

Dr. Jutta Schwarzkopf\\
Leibniz-Institut f\"ur Kristallz\"uchtung (IKZ), 12489 Berlin, Germany\\

\end{affiliations}

\setcounter{figure}{0}
\renewcommand{\figurename}{Figure S}
\setcounter{table}{0}
\renewcommand{\tablename}{Table S}
\newpage

\begin{table}[H]

        \caption{Deposition time and thickness data for different Hf sputter settings.}   
        \centering
    \begin{tabular}{c|cc|ccc}
       
 \multicolumn{1}{c}{\textbf{}} & \multicolumn{2}{c}{\textbf{Set-1}} & \multicolumn{2}{c}{\textbf{Set-2}} \\
        Hf sputter power [W] & Deposition time [min] & Thickness [nm] & Deposition time [min] & Thickness [nm] \\
        \hline
        0 & 60 & 110 & 180 & 270 \\
        15 & 60 & 121 & 180 & 310 \\
        30 & 60 & 128 & 180 & 347 \\
        45 & 60 & 143 & - & - \\
        \hline
    \end{tabular}

    \label{tab:hf_sputter_data}
\end{table}
\begin{figure}[H]
    \centering
    \includegraphics{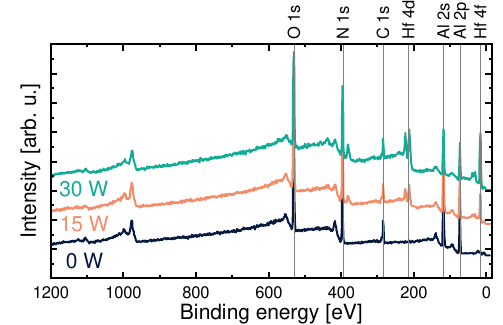}
    \caption{XPS survey taken on films deposited with varying Hf sputter power on $n^+$-Si.
    }
    \label{fig:SI-XPS survey}
\end{figure}
\begin{figure}[H]
    \centering
    \includegraphics{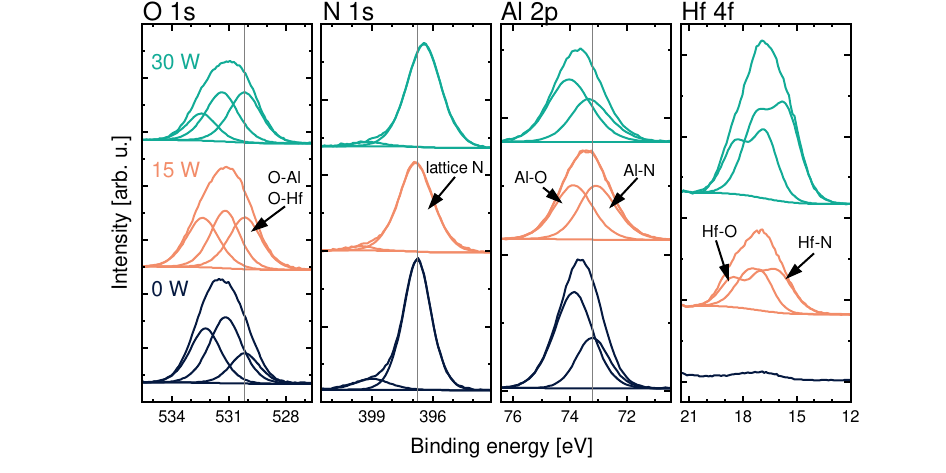}
    \caption{XPS core level region scans with fitting of the compositions taken on films deposited with varying Hf sputter power on $n^+$-Si.
The binding energy was calibrated to adventitious carbon at 284.8\,eV, requiring a shift of 5 -- 7\,eV due to charging of the surface. The films containing Hf displayed a higher degree of charging, indicating no decrease in resistivity with Hf incorporation.}
    \label{fig:SI-XPS_CL}
\end{figure}

All aluminum Al 2p peaks were fitted as one or more pairs of spin–orbit split subpeaks with a separation of 0.4 eV between the Al 2p$_{3/2}$ and Al 2p$_{1/2}$ components. The ratio of the area of the 2p$_{3/2}$ component to the area of the 2p$_{1/2}$ component was fixed at 2 : 1 \cite{Rosenberger.2008}. Similarly, we summarized the two doublets used to fit the Hf 4f region by displaying the sum of the two peaks as one line.\\

\begin{figure}[H]
    \centering

        \includegraphics{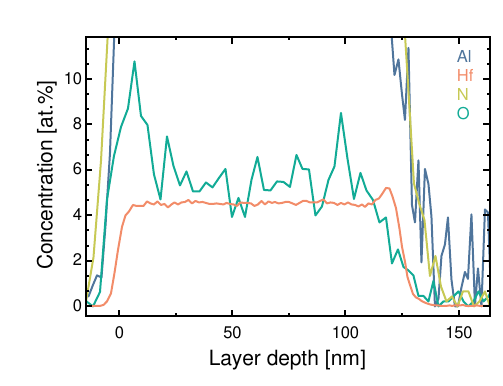}
    \caption{Zoom-in of the ERDA elemental composition depth profile measured on a sample sputtered with 30 W applied to the Hf sputter target.}
    \label{fig:SI_ERDAHF}
\end{figure}
\begin{figure}[H]
    \centering
\includegraphics[]{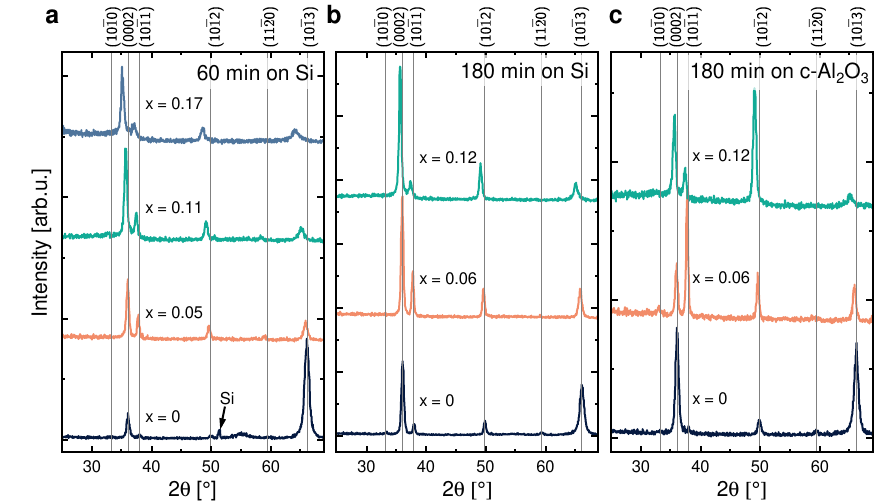}
    \caption{Structure of sputtered wurtzite \Hfx.
   GIXRD-scan of  $\Hfx$ thin films \textbf{a)} deposited for 60\,min on (100)-Si,  \textbf{b)} deposited for 180\,min on (100)-Si, and \textbf{c)} on c-\sap. The gray lines represent the diffraction pattern of polycrystalline wurtzite AlN.
    }
    \label{fig:SI-GIXRD-lattPar}
\end{figure}
\begin{figure}[H]
    \centering
\includegraphics[]{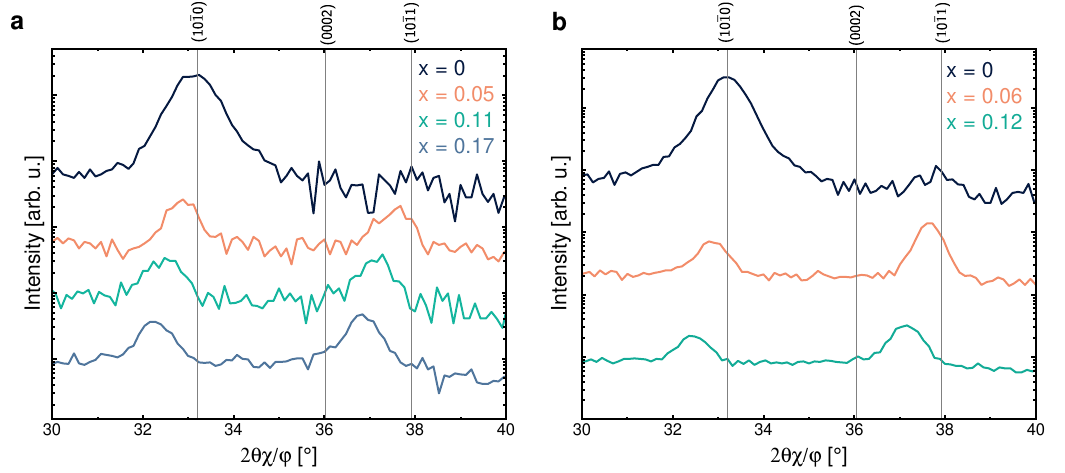}
    \caption{In-plane $2\theta _\chi / \phi$ scans
    of \ce{Al_{1-x}Hf_{x}N} thin films deposited \textbf{a)} on (100)-Si for 60\,min and \textbf{b)} on $c$-\sap~for 180\,min displayed on a logarithmic intensity scale. The vertical gray reference lines indicate the diffraction pattern of wurtzite AlN labeled with their Miller-Bravais indices \cite{Ott.1924}. We note that the in-plane measurements required the use of a non-monochromatized X-ray source, resulting in broadening of all peaks.}
    \label{fig:SI-IPXRD}
\end{figure}

\begin{figure}[H]
    \centering
    \includegraphics{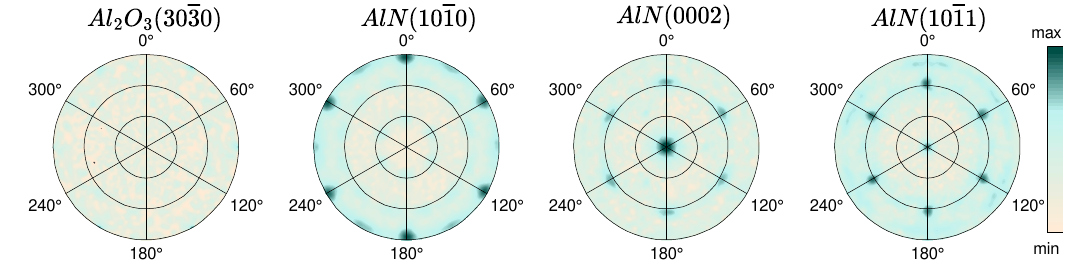}
        \includegraphics{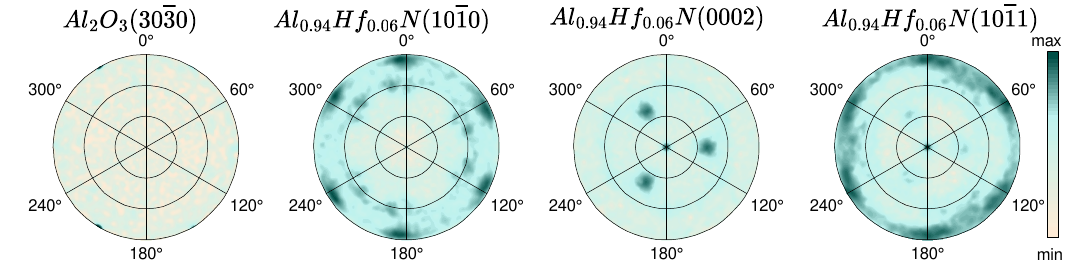}
            \includegraphics{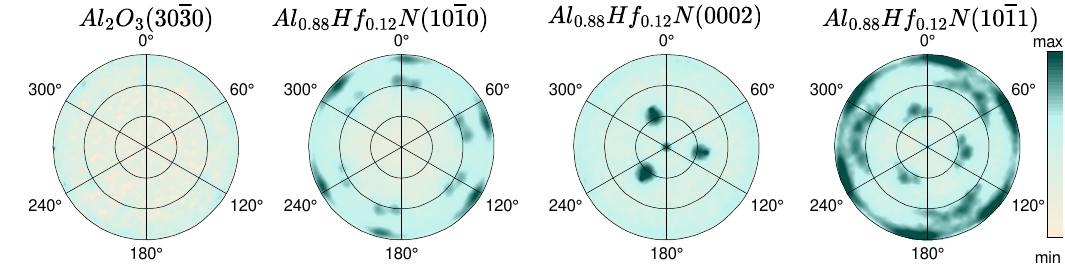}
    \caption{In-plane pole figure measurements on \sap~$(30\overline{3}0)$, AlN $(10\overline{1}0)$, AlN $(0002)$, and AlN $(10\overline{1}1)$ (left to right) measured on sputtered, undoped AlN,  \ce{Al_{0.94}Hf_{0.06}N}, and  \ce{Al_{0.88}Hf_{0.12}N} (top to bottom). All pole figures are on an individual log scale. }
    \label{fig:SI_IPPFsummary}
\end{figure}

\threesubsection{TEM analysis of semipolar epitaxial relations}

\begin{figure}[H]
    \centering
    \includegraphics[width=0.8\linewidth]{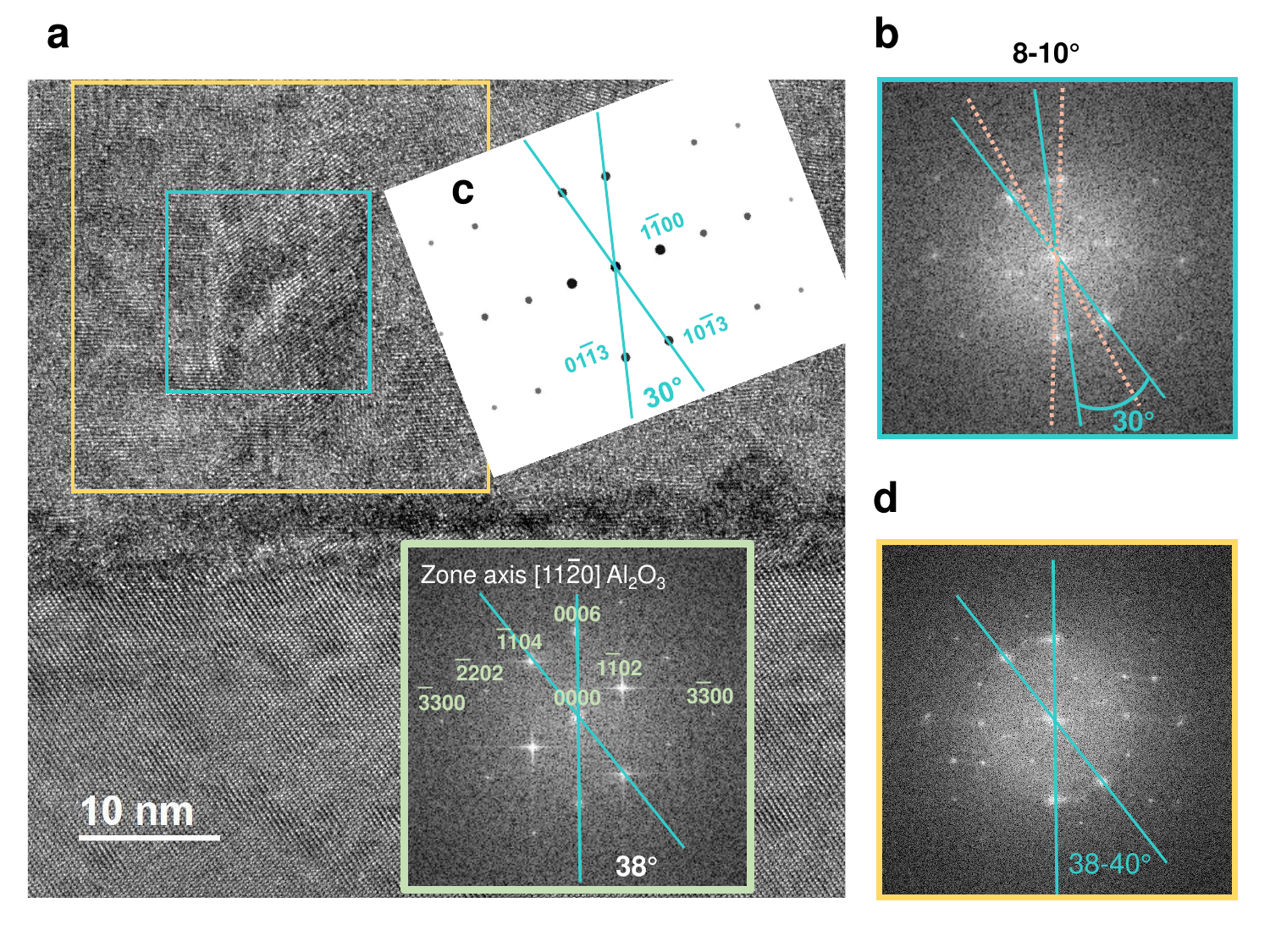}
    \caption{\textbf{a)} HRTEM image of {\Hf{0.12}{0.88}} grown on $c$-\sap~along the \sap~[$11\bar20$] zone axis (substrate FFT pattern in light green box), illuminating the growth condition for $\{10\bar13\}$. \textbf{b)} FFT pattern of the area framed by cyan box in a. The observed spots form an angle of 30$^\circ$. \textbf{c)} AlN [$11\bar21$] zone axis calculated using JEMS program.    \textbf{d)} FFT pattern obtained from the area marked in yellow box in a, showing the presence of distributed $\{10\bar13\}$ growth conforming to the \sap~lattice.}
    \label{fig:TEM2}
\end{figure}

We display an HRTEM image taken along \sap~~zone axis, [$11\bar20$] in Figure~\ref{fig:TEM2}a. The FFT pattern of the substrate can be seen in the inset framed with a light green box, where we indicate the $38^\circ$ angle between $(\bar1104)$ and $(0006)$. Figure~\ref{fig:TEM2}b corresponds to the FFT pattern of the area framed with a cyan box in a, in which we identify two identical zone axes offset by $8\sim10^\circ$. The observed reflections each form an angle of $30^\circ$. Figure~\ref{fig:TEM2}c displays the zone axis calculated for AlN [$11\bar21$] using the JEMS program \cite{JEMS}. There is a striking resemblance between Figure~\ref{fig:TEM2}b, c, and the $c$-\sap~pattern. The FFT in Figure~\ref{fig:TEM2}d is obtained from the area marked by a yellow box in Figure~\ref{fig:TEM2}a. It displays a commonly observed FFT of the \Hfx~layer when inspected along the \sap~zone axis [$11\bar20$], where the same crossed pattern is slightly spread out, covering an angle of $38\sim40^\circ$, which coincides with the \sap~pattern of $38^\circ$. Evidently, \Hf{0.12}{0.88} attempts to conform to the $(\bar1104)$ - $(0006)$ structure in the $c$-\sap~substrate by following the $\{10\bar13\}$ orientation. This explains the appearance of $(10\bar13)$ in the XRD (Figure~3a. Although $c$-\sap~is treated as hexagonal, it is in fact rhombohedral. This lower degree of symmetry results in the epitaxial growth of the semipolar plane.

\begin{figure}[H]
    \centering
    \includegraphics{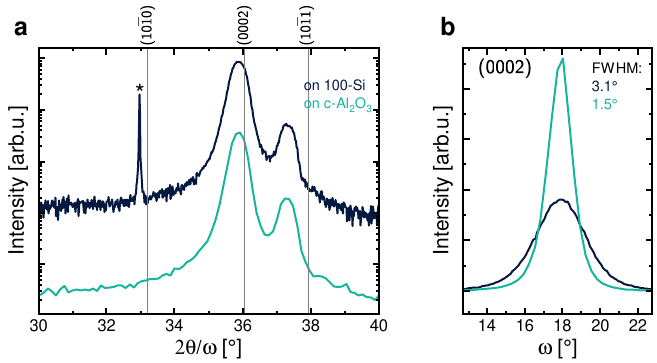}
    \caption{Structural characterization of sputtered Al\textsubscript{0.9}Hf\textsubscript{0.1}N thin films grown with an AlN buffer layer. (a) HRXRD 2$\theta/\omega$ scans of Al\textsubscript{0.9}Hf\textsubscript{0.1}N thin films deposited on c-Al\textsubscript{2}O\textsubscript{3} and (100) Si following growth of an AlN buffer layer, displayed on a logarithmic intensity scale. The vertical gray reference lines indicate the reference diffraction positions of wurtzite AlN, labeled with their Miller-Bravais indices. (b) Rocking curve scans on the (0002) reflection of the respective wurtzite Al\textsubscript{0.9}Hf\textsubscript{0.1}N films deposited on top of AlN buffer layers.  }
    \label{fig:AlN40}
\end{figure}
\begin{figure}[H]
    \centering
    \includegraphics{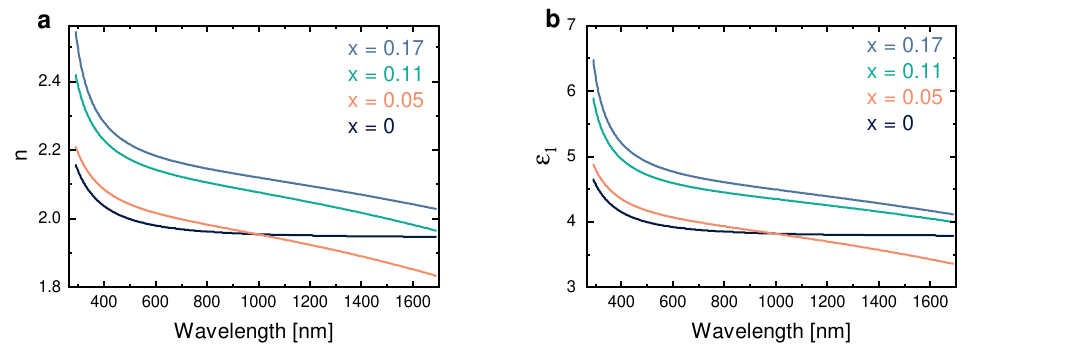}
    \caption{Optical constants of $\Hfx$ on Si.
    \textbf{a)} Refractive index, $n$, and \textbf{b)} dielectric function, $\epsilon_1$, as a function of wavelength determined by spectroscopic ellipsometry.}
    \label{fig:SI_SEresults}
\end{figure}

\begin{figure}[H]
    \centering
    \includegraphics{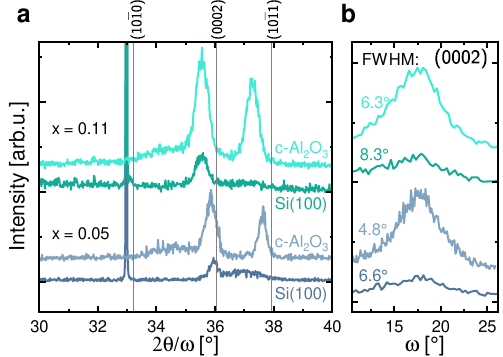}
    \caption{Structure of sputtered of wurtzite \ce{Al_{1-x}Hf_{x}N} on (100)-Si and c-\sap. 
    \textbf{a)} $2\theta/\omega$- scan of $\Hfx$ thin films with \textbf{b)} XRC-scans on the peak assigned with $(0002)$ in wurtzite AlN and labeled FWHM.}
    \label{fig:SI-HRXRDsubstrate}
\end{figure}

\begin{figure}[H]
    \centering
    \includegraphics[width=14cm]{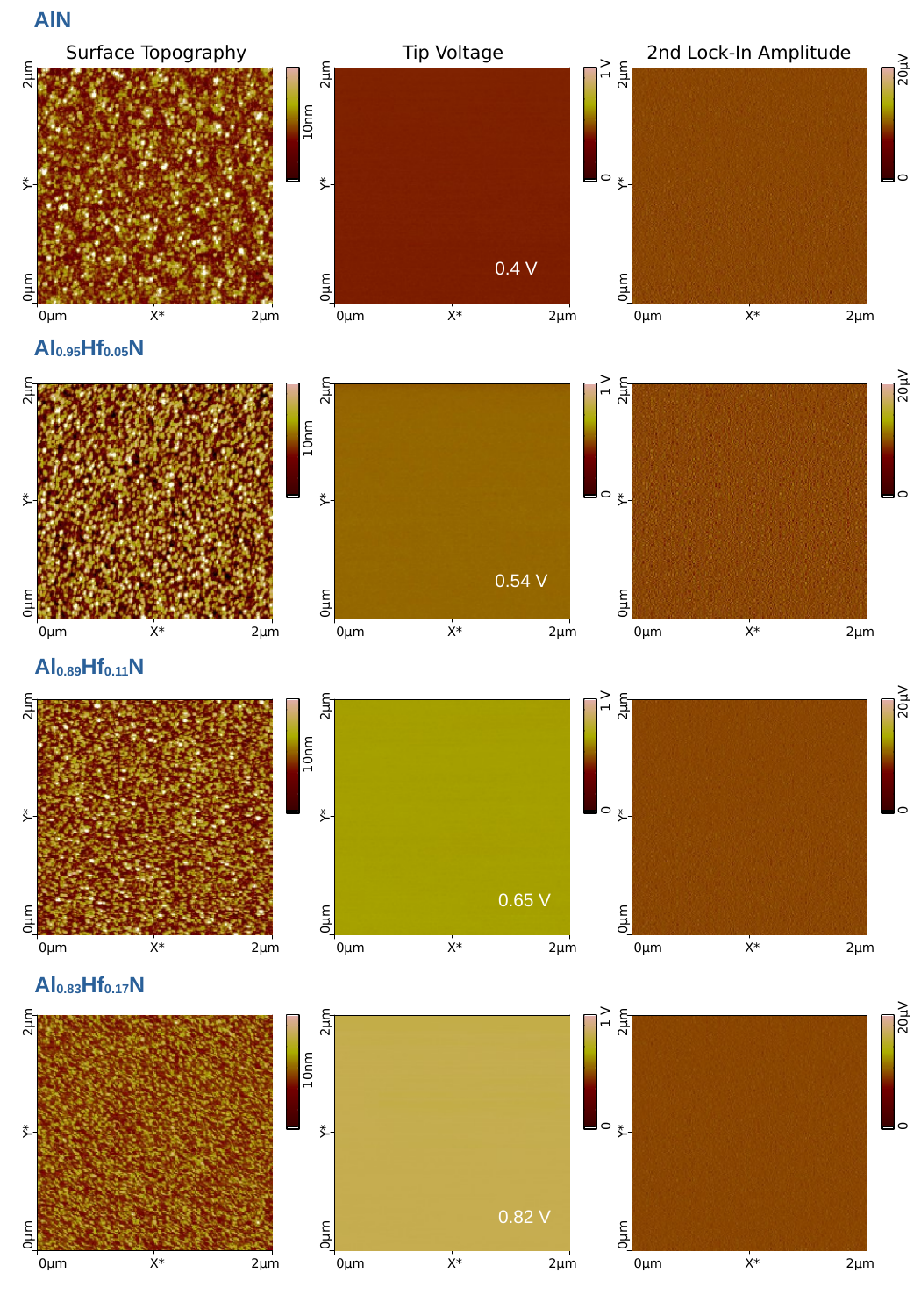}
    \caption{KPFM scans of Al\textsubscript{1-x}Hf\textsubscript{x}N thin films (60 min deposition series).  Left: surface topography; middle: tip voltage; right: secondary lock-in amplitude. The tip voltage is dynamically adjusted to minimize the secondary lock-in amplitude and is used to determine the local surface potential. The measurements show a spatially uniform surface potential across the scanned areas, and the area-averaged tip voltage is indicated on each map. A small instrumental offset of 10 $\mu$V is visible in the lock-in amplitude maps. }
    \label{fig:SI-KPFM}
\end{figure}

\begin{figure}[H]
    \centering
    \includegraphics[width=14cm]{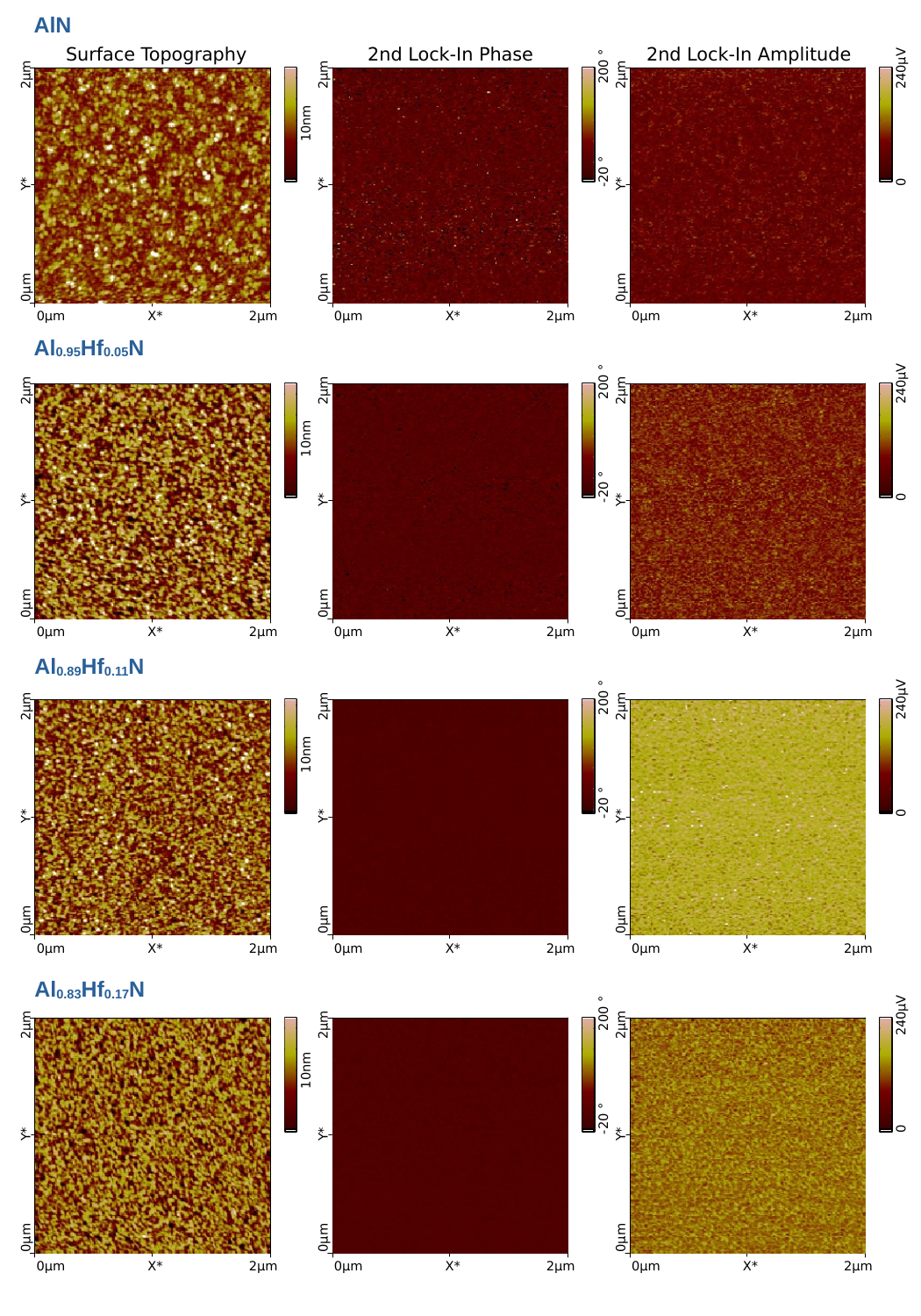}
    \caption{PFM scans of Al\textsubscript{1-x}Hf\textsubscript{x}N thin films (60 min deposition series). Left: surface topography; middle: secondary lock-in phase; right: secondary lock-in amplitude. During the measurements, the tip-sample potential difference was compensated using the tip voltage determined from the KPFM measurements. The area-averaged lock-in amplitude was used to calculate the corresponding \textit{d}\textsubscript{33} values shown in Figure 6, while the standard deviation across the scan regions was used to determine the error bars.  }
    \label{fig:SI-PFM}
\end{figure}
\begin{figure}[H]
    \centering
    \includegraphics{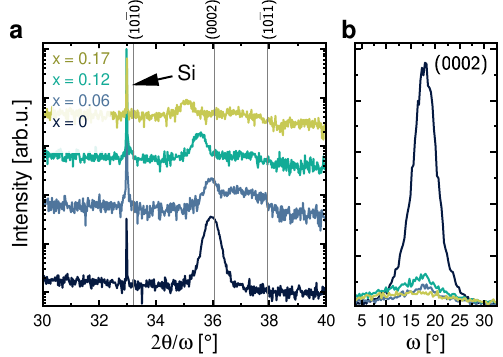}
    \caption{Structure of sputtered of wurtzite \ce{Al_{1-x}Hf_{x}N} on (100)-Si. 
    \textbf{a)} $2\theta/\omega$- scan of $\Hfx$ thin films with \textbf{b)} XRC-scans on the peak assigned with $(0002)$ in wurtzite AlN.}
    \label{fig:SI-HRXRDset1}
\end{figure}
\begin{figure}[H]
    \centering
    \includegraphics{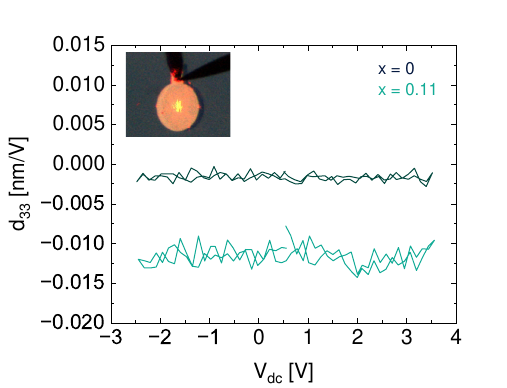}
    \caption{Piezoelectric strain coefficient obtained by DBLI measurements as a function of dc bias, $V_{dc}$, for AlN (x = 0) and \Hf{0.11}{0.89} ~(x = 0.11). The bias was cycled between -2.5 and 3.5 V, starting in the positive direction. The inset shows the 400 µm-diameter top electrode. }
    \label{fig:SI-DBLI}
\end{figure}

\threesubsection{Evaluation of $\ktwo$}

To quantify the SAW excitation performance, we extracted the electromechanical coupling coefficient, $\ktwo$, by fitting the experimental data to the modified Butterworth-Van Dyck (BvD) circuit model and evaluating the admittance \cite{Smith_ITMTT17_856_69}. 
The corresponding results for the data presented in Figure~7a are summarized in Table \ref{k2}, together with values extracted from the resonance/anti-resonance read-off method \cite{Hashimoto_Book_09}, commonly used for thin-film bulk acoustic resonators (fBARs). Both approaches confirm enhanced $\ktwo$ values for the Hf-containing films. In particular, the binary AlN reference film exhibits $\ktwo<0.01$, while \ce{Al_{0.94}Hf_{0.06}N} reaches a value of $\ktwo=0.03$. Importantly, both resonators maintain high quality factors of $Q \approx 300$ (Table S\ref{k2}), resulting in a superior figure of merit, $\ktwo\times Q$, for the Hf-containing film. Although the absolute values of these parameters remain smaller than those of bulk AlN, they agree with expectations for the large electrode-to-film thickness ratio used here. Comparable signal levels have been reported for the Rayleigh (R) mode in \ce{Al_{0.91}Sc_{0.09}N} films thicker than $2~\mu$m and operating below 1~GHz \cite{Thierry2019}.
We anticipate that the coupling can be further increased through improved growth of epitaxial films, as well as by engineering devices with thicker films and thinner or heavier metal electrodes \cite{Tang_2016}. 

At $x = 0.12$ $\ktwo$ returns to a level similar to x = 0. An increase in effective $\epsilon_{33}$ from 13 for x = 0.06 to 16 for x = 0.12 (Table S~\ref{k2}) would lead to an approximately 20\% reduction in $\ktwo$. This change is significantly smaller than the experimentally observed reduction in $\ktwo$ from 0.03 to 0.007 between these two compositions, indicating that the degradation in electromechanical coupling at $x = 0.12$ is dominated by increased structural disorder rather than by the dielectric response alone. This interpretation is also consistent with the more than threefold enhancement in $\ktwo$ observed for $x = 0.06$ relative to AlN, despite only a minor change in the dielectric constant.

\begin{table}[h!]
 \caption{Electromechanical coupling coefficients and Q-factors for $\lSAW=1$ $\mu$m.}
 \centering
  \begin{tabular}[htbp]{@{}llll@{}}
    \hline
    $x$ in \Hf{x}{1-x} & $\ktwo$ (\%), read-off & $\ktwo$ (\%), fitted & $Q$, fitted\\
    \hline
    0  & 0.7  & 0.008  & 345\\
    0.06  & 1.1  & 0.03  & 328\\
    0.12 & 0.6 & 0.007 & 511 \\
    \hline
  \end{tabular}
  \label{k2}
\end{table}

\begin{figure}[H]
    \centering
    \includegraphics{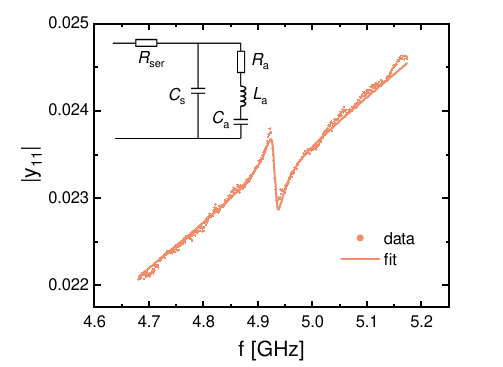}
    \caption{Modified BvD fit for the acoustic resonance at 4.9 GHz, $\lSAW=1$~$\mu$m, \Hf{0.06}{0.94}}
    \label{fig:yfit}
\end{figure}

For the modified BvD method, we fit the y-parameters of the experimental $\Sone$ resonances to the circuit model illustrated in the inset of Figure~S\ref{fig:yfit}. Included is a series resistor $R_\mathrm{ser}$ determined by the peripheral circuit. The acoustic branch is modelled by the acoustic resistor $R_\mathrm{a}$, acoustic capacitor $C_\mathrm{a}$ and acoustic inductor $L_\mathrm{a}$. The capacitor in the extra branch $C_\mathrm{s}$ is determined by the geometric capacitor of the IDT. An example of the fitted result corresponding to \Hf{0.06}{0.94} in Figure 7a is shown in Figure~S\ref{fig:yfit}. Within this approach, $\ktwo$ is obtained according to \cite{Smith_ITMTT17_856_69}: 
\begin{equation}
    \ktwo=\frac{\pi}{4N}\frac{G_0}{B_0},
\end{equation}
where $G_0$ is the acoustic conductance and $B_0$ the total reactance, both evaluated at the acoustic resonance.

\begin{figure}[H]
    \centering
    \includegraphics{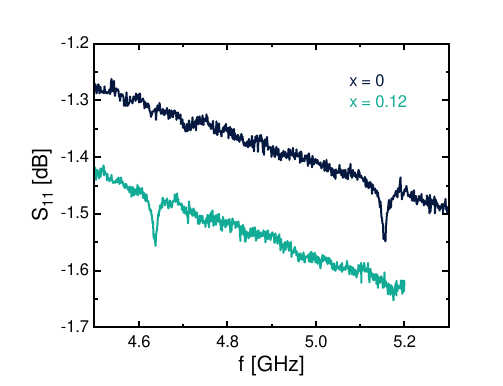}
    \caption{$\Sone$ for IDTs of $\lSAW=1$ $\mu$m along \Hfx/\sap~$\langle1\bar100\rangle$/$\langle11\bar20\rangle$, AlN vs. \Hf{0.12}{0.88}. }
    \label{fig:x12}
\end{figure}

\threesubsection{Refinement of stiffness tensor elements}

For the 6mm point group with out-of-plane c-axis, the velocity of the Rayleigh mode depends on the uniaxial $C_{11}$, $C_{33}$, and the shear element $C_{55}$ ($= C_{44}$), whereas the shear-horizontal mode depends strongly on $C_{66}=(C_{11}-C_{12})/2$ and also on $C_{44}$ through the film/substrate boundary conditions. Therefore, the refinement was restricted to $C_{11}$, $C_{12}$, $C_{33}$, and $C_{44}$, while maintaining the mechanical stability conditions of the hexagonal structure and leaving the remaining parameters unchanged. The initial parameters for AlN were taken from literature values, while those for \Hf{0.12}{0.88} ~were taken from previously published DFT calculations. Since corresponding stiffness coefficients are not available for $x = 0.06$, the initial values for this composition were interpolated between those for AlN and $x = 0.12$. 

For AlN, $C_{11}$, $C_{33}$, and $C_{44}$ were first scaled together by a factor of 0.9 to account for the lower measured acoustic velocities. Subsequently, $C_{44}$ was further refined to improve agreement with the experimentally observed dispersion. For \Hf{0.06}{0.94}, no independent theoretically calculated stiffness tensor was available. Given the relatively small Hf content, we instead used the refined tensor elements from AlN as a starting point for subsequent refinement of the Hf-containing film. Here, $C_{11}$ was held constant, while $C_{12}$ was increased to reduce the value of $C_{66}$ and account for the lower shear-horizontal mode velocities recorded experimentally. $C_{33}$ and $C_{44}$ were subsequently refined to improve agreement with the experimental data for the Rayleigh mode.

\begin{table}[H]
\begin{center}
\caption{Initially used stiffness $C_{IJ}$, piezoelectric stress $e_{iJ}$, and dielectric $\epsilon_{ij}$ constants, as well as densities $\rho$ for numerical calculations.}
\begin{tabular}{@{}lllllll}
\hline
Modulus&AlN&\Hf{0.94}{0.06} [extrapolated]&Al$_{0.88}$Hf$_{0.12}$N&Ti&Al\cite{Slobodnik73a}&c-\sap\cite{Royer00a}\\
\hline
~$C_{IJ}$ [GPa]&\cite{wang_piezoelectric_2024}&&\cite{wang_piezoelectric_2024}&[computed]&&\\
$C_{11}$&380&355&330&169&111.3&497\\
$C_{12}$&125&135&140&80.7&59.1&163\\
$C_{13}$ \cite{Deger98a}&100&&100&80.7&59.1&111\\
$C_{14}$&&&&&&-23.5\\
$C_{33}$&354&317&280&169&111.3&498\\
$C_{44}$&112&102&93&44\cite{Ti}&26.1&147\\
\hline
$e_{iJ}$ [C/m$^2$] &&&&&&\\
$e_{15}$ \cite{Tsubouchi_IEEESU32_634_85}&-0.48&-0.48[AlN]&-0.48[AlN]&&&\\
$e_{33}$&1.45&1.78&2.1&&&\\
$e_{31}$ \cite{Tsubouchi_IEEESU32_634_85}&-0.58&-0.58[AlN]]&-0.58[AlN]&&&\\
\hline
$\epsilon_{ij}$, effective &&&&&&\\
$\epsilon_{33}$ [experiment]&13.5&13&16&&&\\
\hline
$\rho$ [kg/m$^3$]&3230\cite{Goldberg_01}&3930 [computed]&4580 [computed]&4506\cite{CRCHandbook}&2700&3986\\
\label{Cij}
\end{tabular}
\end{center}
\end{table}

\begin{table}[H]
\begin{center}
\caption{Tuned stiffness $C_{IJ}$ constants and density $\rho$ for refined numerical calculations. Blank entries were unchanged with respect to the initial parameters listed in Table S\ref{Cij}}
\begin{tabular}{@{}llll}
\hline
Modulus&AlN&\Hf{0.94}{0.06}&Al$_{0.88}$Hf$_{0.12}$N\\
\hline
$C_{11}$&335&335&\\
$C_{12}$&125&160&\\
$C_{33}$&310&330&\\
$C_{44}$&85&98&\\
\hline
$\rho$&&&4400\\
\label{Ctuned}
\end{tabular}
\end{center}
\end{table}

\threesubsection{FEM simulation of SAW and BAW}

\begin{figure}[H]
    \centering
    \includegraphics[width=0.7\linewidth]{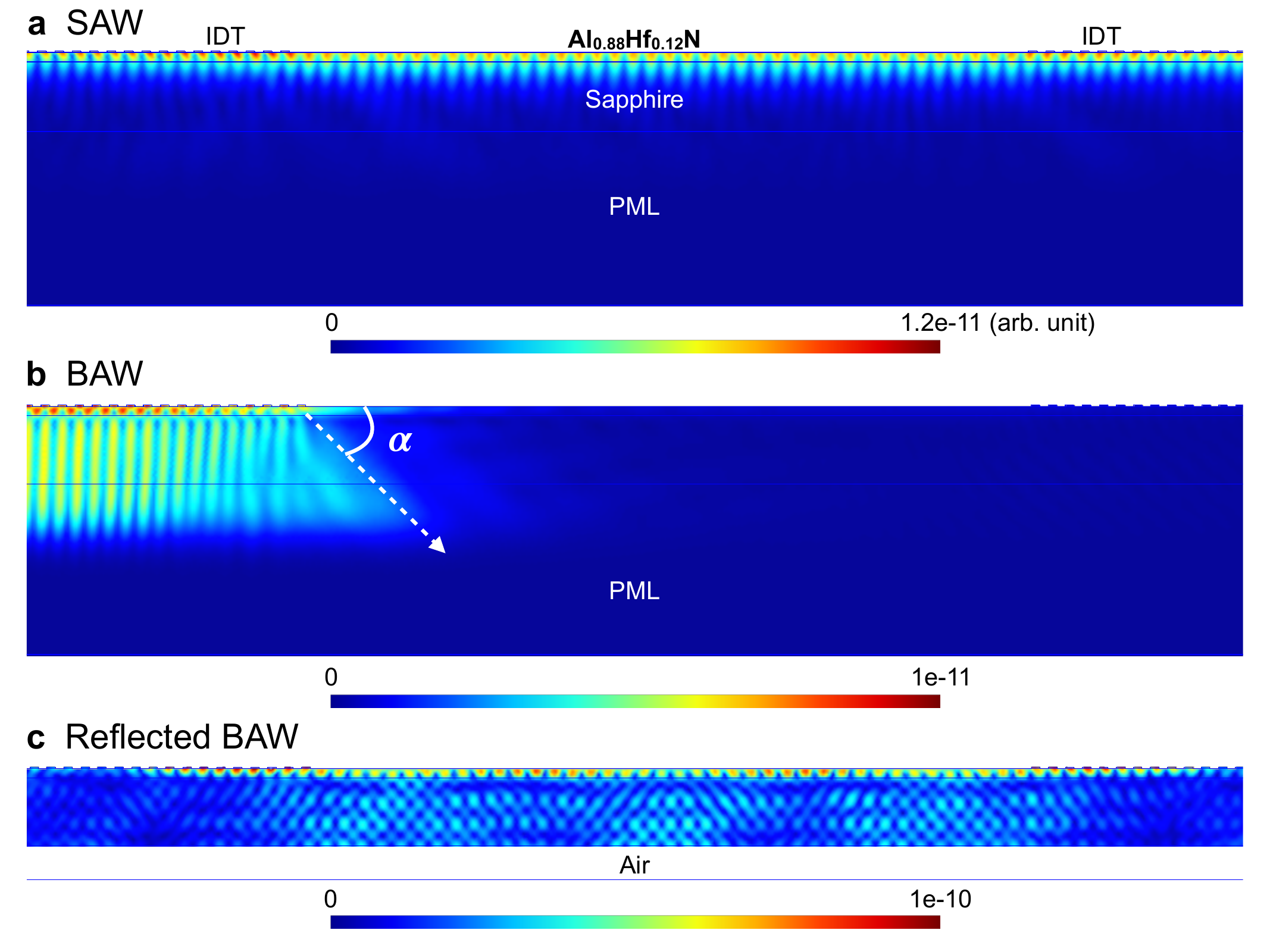}
    \caption{Finite-element simulation of the SAW and BAW modes (partial view). {\textbf{a)}} R-SAW propagating along the surface. \textbf{b)} BAW penetrating into the substrate and absorbed by the PML. \textbf{c)} BAW reflected from the bottom of the substrate (substrate/air interface), reaching the other IDT.}
    \label{fig:FEM}
\end{figure}

To distinguish the nature of the SAWs and BAWs, we perform a simulation with the finite-element method (FEM) using programs GetDP \cite{getdp} and Gmsh \cite{gmsh}. We plot the total displacement in Figure S\ref{fig:FEM} (showing only part of the IDT fingers) for \Hf{0.12}{0.88}, $\lSAW=1.5$ $\mu$m, using parameters from Table S\ref{Cij}. Here, the electrode and film thicknesses are kept true to the experiment, whereas the $c$-\sap~thickness, $N$ and $d$ are reduced, and a perfectly matched layer (PML) is added. At 3.5 GHz (Figure S\ref{fig:FEM}a, the wave is confined close to the surface and propagates from the emitting IDT to the receiving IDT, as expected for the Rayleigh (R) SAW mode. In Figure S\ref{fig:FEM}b, plotted for 5.3 GHz, instead of propagating towards the receiving IDT, the wave penetrates deep into the substrate and the PML, with an angle $\alpha$ visible at the edge of the emitting IDT, indicating a BAW mode. In Figure S\ref{fig:FEM}c we replace the PML with air, simulating the polished backside of the $c$-\sap~substrate. Without the absorption from the PML the BAW is reflected at the $c$-\sap-air interface and the reflected wave can reach the receiving IDT. As a result of the much reduced $c$-\sap~thickness $h_\mathrm{s}$ the BAW is reflected three times in the plot. In the experiment there is one reflection.

\bibliographystyle{MSP}